\documentstyle[psfig]{mn}

\newif\ifAMStwofonts

\ifoldfss
  \ifCUPmtlplainloaded \else
    \NewTextAlphabet{textbfit} {cmbxti10} {}
    \NewTextAlphabet{textbfss} {cmssbx10} {}
    \NewMathAlphabet{mathbfit} {cmbxti10} {} 
    \NewMathAlphabet{mathbfss} {cmssbx10} {} 
  \fi
  \ifAMStwofonts
    \ifCUPmtlplainloaded \else
      \NewSymbolFont{upmath} {eurm10}
      \NewSymbolFont{AMSa} {msam10}
      \NewMathSymbol{\upi}     {0}{upmath}{19}
      \NewMathSymbol{\umu}     {0}{upmath}{16}
      \NewMathSymbol{\upartial}{0}{upmath}{40}
      \NewMathSymbol{\leqslant}{3}{AMSa}{36}
      \NewMathSymbol{\geqslant}{3}{AMSa}{3E}

      \let\leq=\leqslant 
       
    \fi
  \fi
\fi 

\ifnfssone
  \newmathalphabet{\mathit}
  \addtoversion{normal}{\mathit}{cmr}{m}{it}
  \addtoversion{bold}{\mathit}{cmr}{bx}{it}
  \newmathalphabet{\mathbfit} 
  \addtoversion{normal}{\mathbfit}{cmr}{bx}{it}
  \addtoversion{bold}{\mathbfit}{cmr}{bx}{it}
  \newmathalphabet{\mathbfss} 
  \addtoversion{normal}{\mathbfss}{cmss}{bx}{n}
  \addtoversion{bold}{\mathbfss}{cmss}{bx}{n}
  \ifAMStwofonts
    \ifCUPmtlplainloaded \else
      \UseAMStwoboldmath
      \makeatletter
      \new@mathgroup\upmath@group
      \define@mathgroup\mv@normal\upmath@group{eur}{m}{n}
      \define@mathgroup\mv@bold\upmath@group{eur}{b}{n}
      \edef\UPM{\hexnumber\upmath@group}
      \new@mathgroup\amsa@group
      \define@mathgroup\mv@normal\amsa@group{msa}{m}{n}
      \define@mathgroup\mv@bold\amsa@group{msa}{m}{n}
      \edef\AMSa{\hexnumber\amsa@group}
      \makeatother
      \mathchardef\upi="0\UPM19
      \mathchardef\umu="0\UPM16
      \mathchardef\upartial="0\UPM40
      \mathchardef\leqslant="3\AMSa36
      \mathchardef\geqslant="3\AMSa3E

      \let\leq=\leqslant 

    \fi
  \fi
\fi 

\ifnfsstwo
  \DeclareMathAlphabet{\mathbfit}{OT1}{cmr}{bx}{it}
  \SetMathAlphabet\mathbfit{bold}{OT1}{cmr}{bx}{it}
  \DeclareMathAlphabet{\mathbfss}{OT1}{cmss}{bx}{n}
  \SetMathAlphabet\mathbfss{bold}{OT1}{cmss}{bx}{n}
  \ifAMStwofonts
    \ifCUPmtlplainloaded \else
      \DeclareSymbolFont{UPM}{U}{eur}{m}{n}
      \SetSymbolFont{UPM}{bold}{U}{eur}{b}{n}
      \DeclareSymbolFont{AMSa}{U}{msa}{m}{n}
      \DeclareMathSymbol{\upi}{0}{UPM}{"19}
      \DeclareMathSymbol{\umu}{0}{UPM}{"16}
      \DeclareMathSymbol{\upartial}{0}{UPM}{"40}
      \DeclareMathSymbol{\leqslant}{3}{AMSa}{"36}
      \DeclareMathSymbol{\geqslant}{3}{AMSa}{"3E}

      \let\leq=\leqslant 

    \fi
  \fi
\fi 

\ifCUPmtlplainloaded \else
  \ifAMStwofonts \else 
    \def\upi{\pi}
    \def\umu{\mu}
    \def\upartial{\partial}
  \fi
\fi

\title[Starbursts in radio galaxies]
{Starbursts and the triggering of the activity in nearby powerful radio galaxies}
\author[Tadhunter et al.]
       {C. Tadhunter$^{1}$, T.G. Robinson$^{1}$, R.M. Gonz\'alez Delgado$^{2}$, 
       K. Wills$^{1}$, R. Morganti$^{3}$
	\\
$^{1}$Department of Physics and Astronomy, University of Sheffield,  Sheffield, S3 7RH, UK\\ 
$^{2}$Instituto de Astrofisica de Andalucia, Apdto. 3004, 18080 Granada, Spain \\
$^{3}$ASTRON, PO Box 2, 7990 AA Dwingeloo, The Netherlands\\}

\date{}
\def\ltsim{\ifmmode\stackrel{<}{_{\sim}}\else$\stackrel{<}{_{\sim}}$\fi}
\def\gtsim{\ifmmode\stackrel{>}{_{\sim}}\else$\stackrel{>}{_{\sim}}$\fi}
\begin{document}
\maketitle
\begin{abstract}{\large}
We present high quality long-slit spectra for three nearby powerful radio galaxies ­--
3C293, 3C305, PKS1345+12. These were taken with the aim of characterising the young
stellar populations (YSP), and thereby investigating the
evolution of the host galaxies, as well as the events that triggered the activity. Isochrone
spectral
synthesis modelling of the wide wavelength coverage spectra of nuclear and off-nuclear
continuum-emitting regions have been used to estimate the ages, masses and luminosities
of the YSP component, taking full account of reddening effects and potential
contamination by activity-related components. We find that the YSP make a substantial
contribution to the
continuum flux in the off-nuclear regions on a radial scale of 1 ­-- 20~kpc in all three
objects. Moreover, in two objects we find evidence for reddened post-starburst stellar
populations in the near-nuclear regions of the host galaxies. The YSP are relatively old (0.1
­-- 2~Gyr), massive ($10^9 < M_{YSP} < 2\times10^{10}$ M$_\odot$) and make up a large proportion
($\sim$1 ­-- 50\%) of
the total stellar mass in the regions of galaxies sampled by the observations. Overall, these
results are consistent with the idea that AGN activity in some radio galaxies is triggered by
major gas-rich mergers. Therefore, these radio galaxies form part of the subset of early-type
galaxies that is evolving most rapidly in the local universe. Intriguingly, the results also
suggest that the radio jets are triggered relatively late in the merger sequence, and that there is
an evolutionary link between radio galaxies and luminous/ultra-luminous infrared galaxies. 

\end{abstract}
\section{Introduction}
Powerful radio sources are invariably associated with early-type host galaxies and their
luminous emission lines and powerful radio emission make them stand out at high
redshifts. Therefore, they are potentially unique probes of the evolution of massive early-type 
galaxies in the early universe. However, if we are to use them in this way, it is crucial to
understand the link between the quasar/jet activity and evolutionary processes in the host galaxies. 
In this context it is notable that the
evolution of the co-moving number density of powerful radio sources (Dunlop \& Peacock 1990), shows a 
marked
similarity to that of the evolution of the global star formation rate in the universe 
(Madau et al. 1996). This 
suggests that the triggering of the quasar/jet activity is intimately linked to the evolution of
the general galaxy population. 

Observations of powerful radio galaxies in the local universe provide important clues to the
triggering mechanism for the jet and quasar activity. In particular, the imaging studies of 
Heckman et al. (1986) and
Smith, Heckman \& Illingworth (1989) provide evidence for morphological features such as double nuclei,
arcs, tails and bridges which suggest that, in a substantial subset of radio galaxies, the
AGN activity  is triggered by the accretion of gas during galaxy mergers and interactions.
This is backed up by spectroscopic studies (Tadhunter et al. 1989, Baum et al. 1990)
which show emission line kinematics consistent with an accretion origin for the warm gas 
in many radio galaxies, although a cooling flow origin is a viable alternative to mergers for
some objects at the centres of galaxy clusters.

An attraction of the merger hypothesis for the triggering of the activity is that, potentially, it 
can help to explain the redshift evolution of the radio source population and its similarity to
that of the star formation history of the universe. This is because, in a hierarchical galaxy 
evolution  scenario, galaxy mergers and interaction are not only a major driver for star
formation, but they can also provide the fuel for AGN and jet activity (e.g. Kauffmann \& Haehnelt
2000). Thus, radio galaxies 
may represent a particular, post-merger phase, in the evolution of giant elliptical galaxies.

Despite the promise of the merger hypothesis for the triggering of the activity, the following
uncertainties remain:
\begin{itemize}
\item What types of mergers or galaxy interactions trigger radio jet activity?
\item At what stage during the merger are the jets triggered?
\item Do all massive early-type galaxies go through one or more radio galaxy phase(s) as 
they build up via galaxy mergers?
\item What is the relationship between powerful radio galaxies and other types of merging 
systems in which a massive early-type galaxy is the end point, for example, the ultraluminous infrared 
galaxies (ULIGS)?
\end{itemize}
Unfortunately the quantitative information that can be extracted
from the existing morphological and kinematical data is limited. 
Therefore it is important to
develop alternative techniques for investigating the triggering events. One promising 
alternative is to use
the properties of the starbursts that are expected to be triggered as part of the merger 
process. 

Although a substantial proportion of radio galaxies show blue or UV
excesses compared with normal early-type galaxies (Lilly \& Longair 1984, 
Smith \& Heckman 1989), various activity-related  components can contribute to the optical/UV 
continuum and thereby complicate the interpretation of the continuum spectral energy distributions
(SEDs) in terms of the underlying
stellar populations. Activity-related
continuum components include: scattered AGN light (e.g. Tadhunter et al. 1992, Cimatti et al. 1993,
Cohen et al. 1999), nebular 
continuum from the emission line nebulae (Dickson et al. 1995) and direct light from
moderately extinguished quasar nuclei (Shaw et al. 1995). However, with careful modelling of spectroscopic and
polarimetric data it is possible to accurately quantify the
contributions of these components to the optical/UV continua and thereby study the underlying
stellar populations of the host galaxies. In this way it has been possible to show that 
young stellar populations (YSP) make a significant
contribution to the optical/UV continua  in 25 ­-- 40\% of powerful radio galaxies at low and 
intermediate redshifts ($z < 0.7$: Aretxaga et al. 2002, Tadhunter et al. 2002, Wills et al. 2002, 2004).

A further complicating factor is that, even if young stellar populations can be identified, they are not
necessarily related to the normal evolution of the host galaxies. It has been proposed that the shocks
induced by the radio jets as they sweep through the ISM of the host galaxies may trigger star formation,
and be a major cause of the alignment effect in high 
redshift radio galaxies (e.g. Rees 1989). However, apart from a few well-publicised examples in which there is good
evidence that the excess UV emission along the radio axis is due YSP rather than
scattered light or nebular continuum (e.g. Minkowski's object:
van Breugel et al. 1985), direct observational evidence for jet-induced star formation 
is lacking in the overwhelming majority of radio galaxies.

In order to realise the full potential of radio galaxies as evolutionary probes, it is now important to
establish the detailed properties of their YSP. In this, the first in a series of papers in which
we will investigate the nature of the YSP in radio galaxies, we present the results of the detailed analysis of
the continuum spectra of 
three nearby powerful radio galaxies --- 3C293, 3C305, PKS1345+12. 

We assume the cosmological parameters $H_0 = 75$ km s$^{-1}$ Mpc$^{-1}$
and $q_0 = 0.0$ throughout this paper. 
For these parameters
1.00 arcsecond corresponds to 0.82, 0.75 and 1.98~kpc at the redshifts of 3C293, 3C305 and 
PKS1345+12 respectively.

\section{The sample}

\begin{table}
{\centering \begin{tabular}{lllccc}
            &           &Galactic    &Radio   &Slit     &Exposure                                               \\
 Object     & \( z \)   &\( E(B-V) \)&axis PA &PA        &time(s)                 \\
\hline
 3C293     &\( 0.045 \)&\( 0.017 \) & \( 90 \)& \( 60 \) &2400\\
 3C305     &\( 0.041 \)&\( 0.026 \) & \( 44 \)&\( 240 \) &3600 \\
 PKS1345+12&\( 0.122 \)&\( 0.034 \) &\( 160 \)&\( 160 \) &2400 \\
\end{tabular}\par}

\caption{Basic data and observational details for the objects in the sample. Radio axes in the
case of 3C293 \& 3C305 are determined from the large-scale radio jets, the Galactic reddening estimates are
taken from Schlegel et al. (1998), and the exposure times refer to both red and blue arms of ISIS.}
\end{table}

The three galaxies chosen for this study have the following features in common: 
they all show
previous spectroscopic, photometric and/or far-IR evidence for recent star formation
activity; they all show clear morphological evidence for galaxy mergers/interactions; 
and they all have compact steep spectrum radio cores, with evidence for strong HI absorption.
We further note that all three
objects are in relatively low density galaxy environments: isolated or groups rather than clusters. 
Although they do not represent an unbiased sub-set of the general population of
powerful radio galaxies in the local universe, they comprise a useful sample for 
investigating the links between star formation and the triggering of the
radio activity in galaxy mergers. Some basic information about the sample objects is given in 
Table 1, while a more detailed description of each object is given below.

\begin{figure*}
\begin{tabular}{c}
\psfig{figure=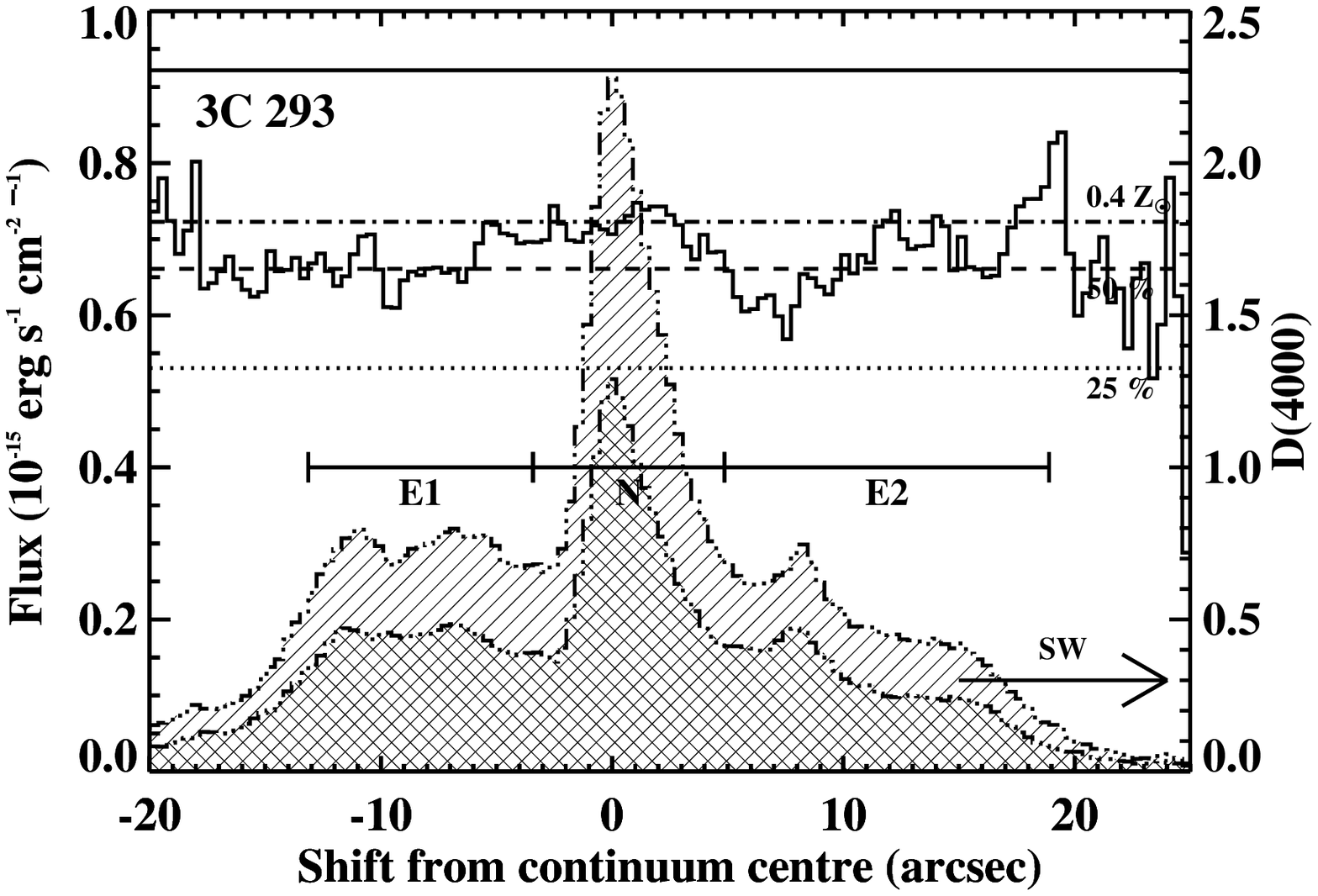,width=9.5cm} \\
\psfig{figure=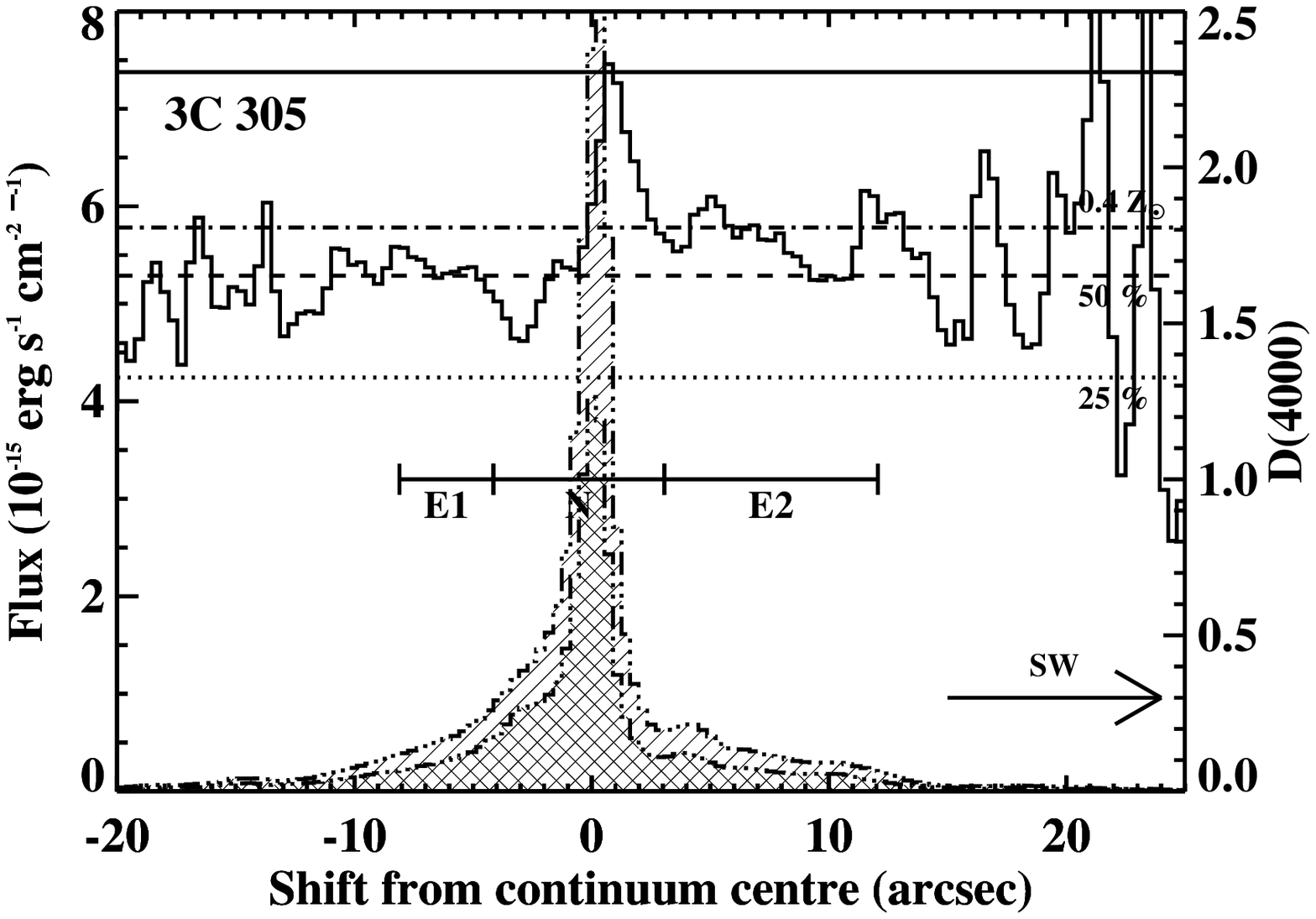,width=9.5cm} \\
\psfig{figure=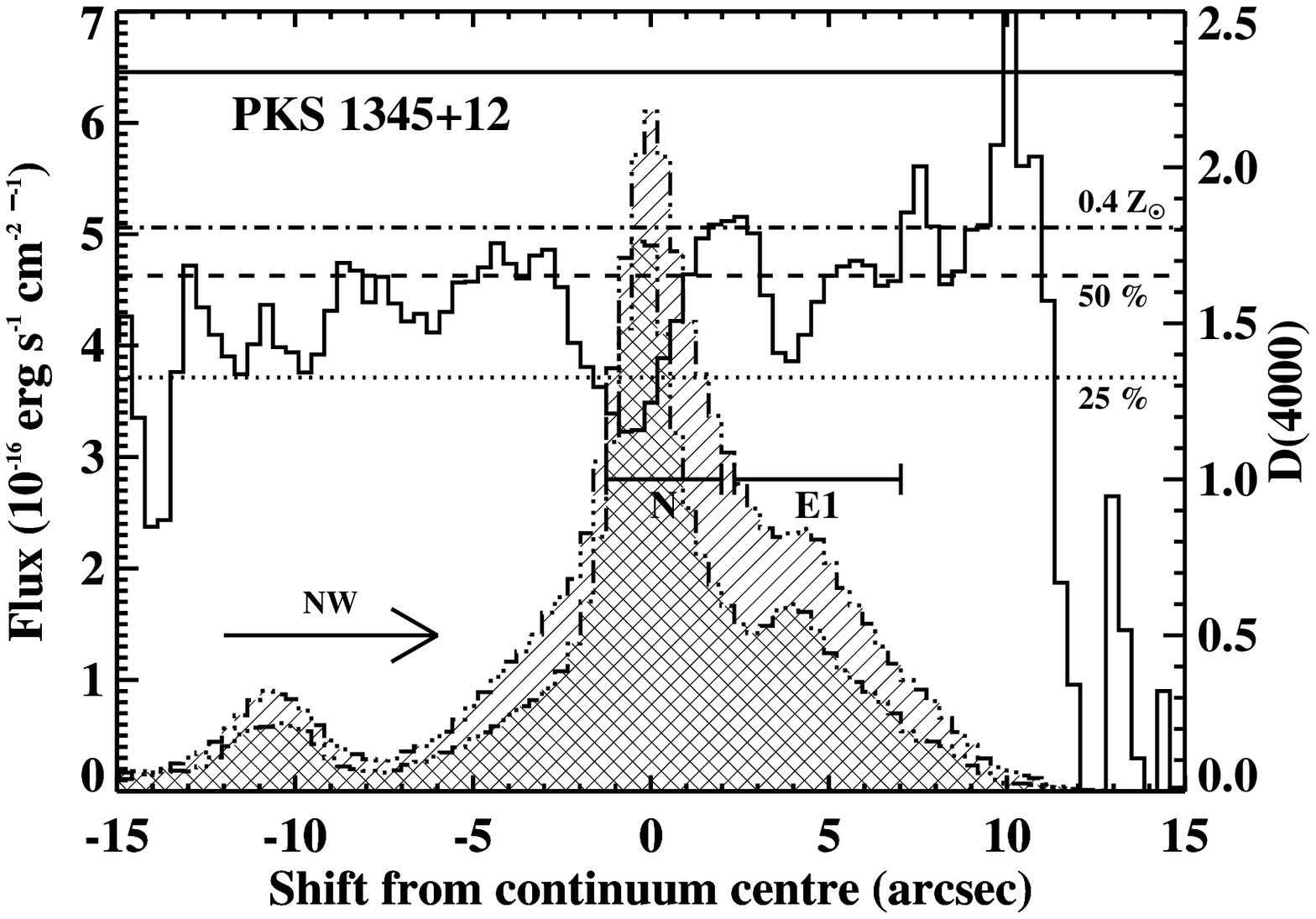,width=9.5cm} \\
\vspace{0.2cm}\\
\end{tabular}

\caption{\label{fig: 4000 break} Spatial profiles of the continuum flux integrated
in the wavelength ranges
\protect\( 3750\protect \) -- \protect\( 3850\protect \) \AA\ (cross hatching) and \protect\( 4150\protect \) --
\protect\( 4250\protect \) \AA\ (hatching)
for 3C 293, 3C 305 \& PKS 1345+12. Also shown is the ratio of the
two (\protect\( D(4000)\protect \), solid line), and the values of
this ratio expected for a \protect\( 12.5\protect \) Gyr elliptical
(solid horizontal line), a \protect\( 12.5\protect \) Gyr elliptical
with \protect\( 50\protect \) \% featureless continuum in the \protect\( 3800\protect \)
\AA\ bin (dashed line) and with \protect\( 75\protect \) \% featureless
continuum (dotted line). The value expected for a sub-solar metallicity
(\protect\( Z=0.4Z_{\odot }\protect \)) elliptical is shown as a
dot-dash line.}
\end{figure*}

A further advantage of the objects in  our sample is that the AGN contribution to the optical/UV
continuum is small in most apertures for most objects. The emission 
line equivalent widths provide an indication of
the contribution of both the nebular continuum (Dickson et al. 1995) and scattered AGN
components (see Tadhunter et al. 2002) to the optical/UV continuum. On the basis of small
emission line equivalent widths we expect the
contribution of AGN-related continuum components to be minor in all spatial
regions for 3C293, in the extended apertures for 3C305, and the in extended aperture for PKS1345+12. 
In the case of the nuclear region of 3C305, the low polarization detected in the
imaging polarimetry observations of Draper et al. (1993) provides evidence that the contribution of
scattered light is relatively small.

\begin{figure*}
\psfig{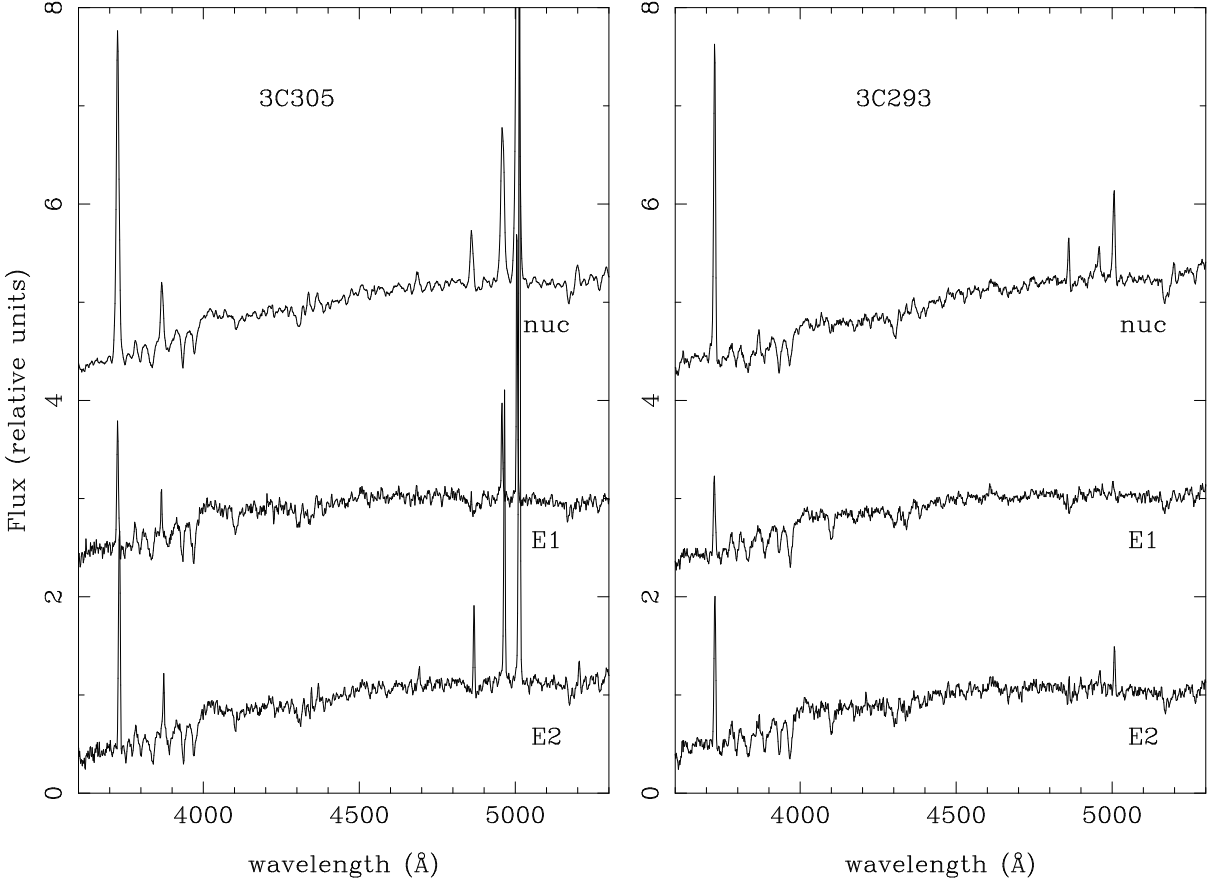}
\caption{Extracted spectra for the extended and nuclear regions in 3C305 and 3C293. See Figure 1 for a
definition of the apertures used to extract the spectra. Only the rest-frame blue/green part of the
spectrum is shown in each case. 
Note that
the age sensitive H$\delta$ line at 4100~\AA, is clearly detected in several of the
apertures.}
\end{figure*}
\begin{figure}
\psfig{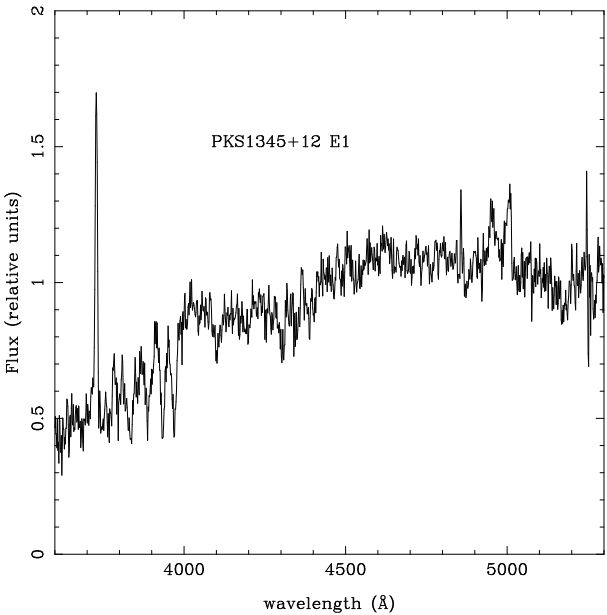}
\caption{Extracted spectrum for the extended region in PKS1345+12. See Figure 1 for a
definition of the aperture used to extract the spectrum. Only the rest-frame blue/green part of the
spectrum is shown.}
\end{figure}

\subsection{3C293}

Although 3C293 ($z = 0.045$) is an FRII radio source with a total extent of more than 160~kpc,
it also shows a compact steep spectrum core, with a complex structure and strong  
HI 21cm absorption (Baan \& Haschik 1981; van Breugel et al. 1984). The impact of the 
activity on the ISM of the host galaxy is manifest in the
broad, blueshifted features detected in both low ionization optical emission lines and HI 21cm
absorption. These latter features provide evidence of  AGN-induced outflows (Morganti et al. 2003).

As a powerful FRII radio galaxy 3C293 is unusual in the sense that it shows a disk-like
morphology in optical images (van Breugel et al. 1984). This, along with the fact that its nuclear regions are crossed by a complex system of 
dust lanes, suggests an S0 morphological classification. However, on a larger scale
the host galaxy shows a jet-like feature linking it to a nearby companion galaxy 28~kpc to the
SW, as well as a fan or tidal tail feature extending 50-60~kpc to the west beyond the 
companion (Heckman et al. 1986). Optical images also show that this object is in an isolated galaxy environment;
there are no galaxies brighter than $M_v = -19.4$ within 130 kpc.
(Smith \& Heckman 1990).

Before the current study, the evidence for recent star formation in this 
object was indirect, comprising: a modest far-IR continuum excess 
($L_{ir} \sim 2.3\times10^{10}$ L$_{\odot}$), bluer (B-V) colours than expected for an
elliptical galaxy at this redshift (Smith \& Heckman 1989), and ultraviolet imaging observations with the HST that
show clumpy structures characteristic of young star clusters (Allen et al. 2002). Perhaps linked to the recent
star formation in this source is the detection of strong, extended CO emission and absorption
close to the core. Making standard assumptions, the total implied mass
of molecular gas is $\sim 1.5\times10^{10}$ M$_{\odot}$ within a radius of 1.5~kpc of the
nucleus of the galaxy (Evans et al. 1999a). 
 
\subsection{3C305}

Unlike 3C293, the radio emission of 3C305 ($z = 0.041$) is relatively compact and shows a distorted ``H'' 
morphology. 
This,  and 
the disturbed kinematics of the optical emission
lines, provide evidence that
the radio jets are interacting strongly with the ISM in the host galaxy (Heckman et al. 1982).

In most other respects 3C305 is similar to 3C293, with a disky optical morphology, complex
dust lanes, evidence for tidal tails (Heckman et al. 1986), blue optical colours (Smith \& Heckman 1989), a modest
far-IR excess ($L_{ir} \sim2.0\times10^{10}$ L$_{\odot}$), an isolated galaxy environment
(Smith \& Heckman 1990),  and evidence for a rich gaseous
environment provided by strong HI 21cm absorption (Jackson et al. 2003). This object had the best
prior evidence
for recent star formation in the sense that Heckman et al. (1982) reported 
the presence of 
Balmer absorption lines characteristic of young stellar populations.

\subsection{PKS1345+12}

The dominant radio emission of PKS1345+12 ($z = 0.122$) is more compact than that in the other
two sources, with total diameter of only 0.15 arcseconds (Stanghellini et al. 2001). Strong HI 21cm absorption is
detected against this compact radio structure, which has a distorted jet-lobe
morphology (Mirabel 1989, Morganti et al. 2004). Recent optical spectroscopy 
provides clear evidence for a powerful outflow
in the reddened, near-nuclear emission line regions of the western
nucleus, perhaps indicating that a strong jet-cloud
interaction is taking place (Holt et al. 2003).

The optical morphology of PKS1345+12 differs from that of the other two objects in the
sense that it shows a double nucleus (separation $\sim$3.6~kpc) surrounded by an
elliptical-like envelope (Axon et al. 2000). As well as the double nucleus, the presence of spectacular tidal
tails extending out to a radius of $\sim$50~kpc provides further evidence that this source has been 
involved in a major galaxy interaction or merger in the recent past (Heckman et al. 1986). Also based
on optical imaging studies, the work of
Zirbel (1997) demonstrates that galaxy environment of this source is isolated -- at the lower end
of the distribution of the $N_{0.5}^{-19}$ environmental richness parameter (see Allington-Smith et al.
1993 for a definition).

Prior to this study, 
the main evidence for recent star formation in this source comprised the imaging study of Surace
et al. (1998) which showed evidence for young star clusters in the extended halo of the system. In addition,
measurements by the IRAS satellite reveal
and the presence of large far-IR continuum
excess ($L_{ir} =1.1\times 10^{12}$~L$_{\odot}$). Indeed, PKS1345+12 is one of the few radio 
galaxies in the local universe that qualifies as 
an ultraluminous infrared galaxy (ULIGs: see Sanders \& Mirabel 1996 for definition). 
In this context it is also notable that 
CO observations demonstrate the presence of a a large reservoir 
($\sim 1.6\times10^{10}$~M$_{\odot}$) of molecular 
gas within $\sim$2.5~kpc of the primary
(western) nucleus (Evans et al.1999b).

\section{Observations and reductions}

The three objects were observed in 2001 May using the ISIS dual-beam
spectrograph on the William Herschel Telescope on La Palma. By using the R300R 
grating on the red arm, and the R300B grating on the blue, 
we were able to cover a large wavelength range at moderate 
spectral resolution ($\sim$7\AA). This large wavelength range provides a ``long
lever''
for the continuum fitting.

The slit was placed along the major axis of optical emission for 3C293 and 3C305,
and along the radio axis for PKS1345+12 (see Table 1 for slit
PAs). A slit width of 1.3 arcseconds was used throughout. In order to avoid problems with 
differential atmospheric refraction, the objects were all
observed at low airmass and/or with the slit aligned close to parallactic angle. 
In the case of PKS1345+12
the slit was centred on the more active, western, nucleus.

The data were reduced using IRAF and the Starlink FIGARO
packages. First bias subtraction was carried out and cosmic rays removed.
The spectra were then wavelength-calibrated, corrected for atmospheric
extinction and flux calibrated, using the standard stars and arc exposures
taken at the same time as the science images. The data were then 
corrected for both S-distortion and tilt, using stars in the same
2-d frame as the object wherever possible. After correction, the residual
offset in the spatial direction between the two ends of the spectra
was estimated to be \( <0.3 \) pixels. The blue spectrum was re-sampled
to give the same spatial pixel scale as the red, and the red and blue
spectra were aligned with each other and shifted back to the rest
frame, using the mean redshift measured from the brighter emission
lines.

The reduction process resulted in two spectra with useful data in the rest wavelength
ranges \( 3314 \)--\( 6400 \) and \( 5891 \)--\( 8200 \)\AA. There is excellent agreement between the flux levels of the blue and red spectra in the
wavelength region of overlap between the spectra. We estimate a relative flux calibration error of $\sim\pm5$\%  across the
full spectral range of the data, based on several wide-slit observations of flux calibration standard stars 
taken during the observing
run.

\section{Results and Analysis}

The analysis of the continuum spectra for the three objects proceeded via three steps: first, using
the 4000\AA\, break to assess the degree of UV excess and investigate the spatial distribution of the
extended UV emission; second, modelling the shapes of the spectral energy
distributions using a combination of stellar and activity-related continuum
components; third, making a more detailed comparison between the data and the models using the CaIIK and Balmer
absorptions lines. Because they were carried out independently, each of these three steps will now be 
described in turn.

\subsection{The 4000\AA\ break}

To gain an impression of the magnitude of the UV excess in the sample objects,
spatial slices of the continuum on each side of the \( 4000 \) \AA\
break in each source were extracted. These are shown in Figure 1,
along with their ratio, \( D(4000) \) --- a measure of the 4000\AA\, break amplitude. 
We use the definition of \( D(4000) \) given in Tadhunter et al. (2002): 
\begin{equation}
D(4000)={\frac{{\int _{4150}^{4250}F_{\lambda }d\lambda }}{{\int _{3750}^{3850}F_{\lambda }d\lambda }}}.
\end{equation}
This form is better suited to active galaxies than the original form of Bruzual(1983),
since it is not subject to contamination by bright
emission lines such as [NeIII]$\lambda$3869. Moreover, D(4000) is relatively
insensitive to the effects of reddening, unlike broad-band colours
(e.g. Smith \& Heckman 1989), and is not sensitive to redshift (K-correction)
effects.

For comparison, Figure 1 also shows the value
of \( D(4000) \) expected for a solar-metallicity \( 12.5 \) Gyr
elliptical template spectrum taken from the 1996 version of the Bruzual \& Charlot
(1993) spectra synthesis models (hereafter BC96: see Charlot et al. 1996 for a discussion). 
As with the template ellipticals used in the
continuum modelling (see below), this template assumes a Salpeter (1955)
initial mass function (IMF) formed in an instantaneous burst, with
solar metallicity stars in the mass range \( 0.1M_{\odot }\leq M\leq 125M_{\odot } \).
Also shown are values calculated by diluting this old elliptical continuum
with a flat spectrum, such that the old elliptical flux comprises
50\% and 25\% of the flux in the wavelength bin centred
at \( 3800 \) \AA, and the value obtained for a sub-solar metallicity
($Z=0.4Z_{\odot } $) \( 12.5 \) Gyr elliptical. This latter clearly
demonstrates that low metallicities are required to produce the same
effect as dilution by a flat spectrum. In fact, studies of normal
ellipticals, using absorption-line indices, show that their metallicities
are generally solar or super-solar (e.g. Kuntschner \& Davies 1998),
and it is therefore difficult to explain the UV excess in terms of
sub-solar abundances.
\begin{figure*}
\begin{tabular}{c}
\psfig{figure=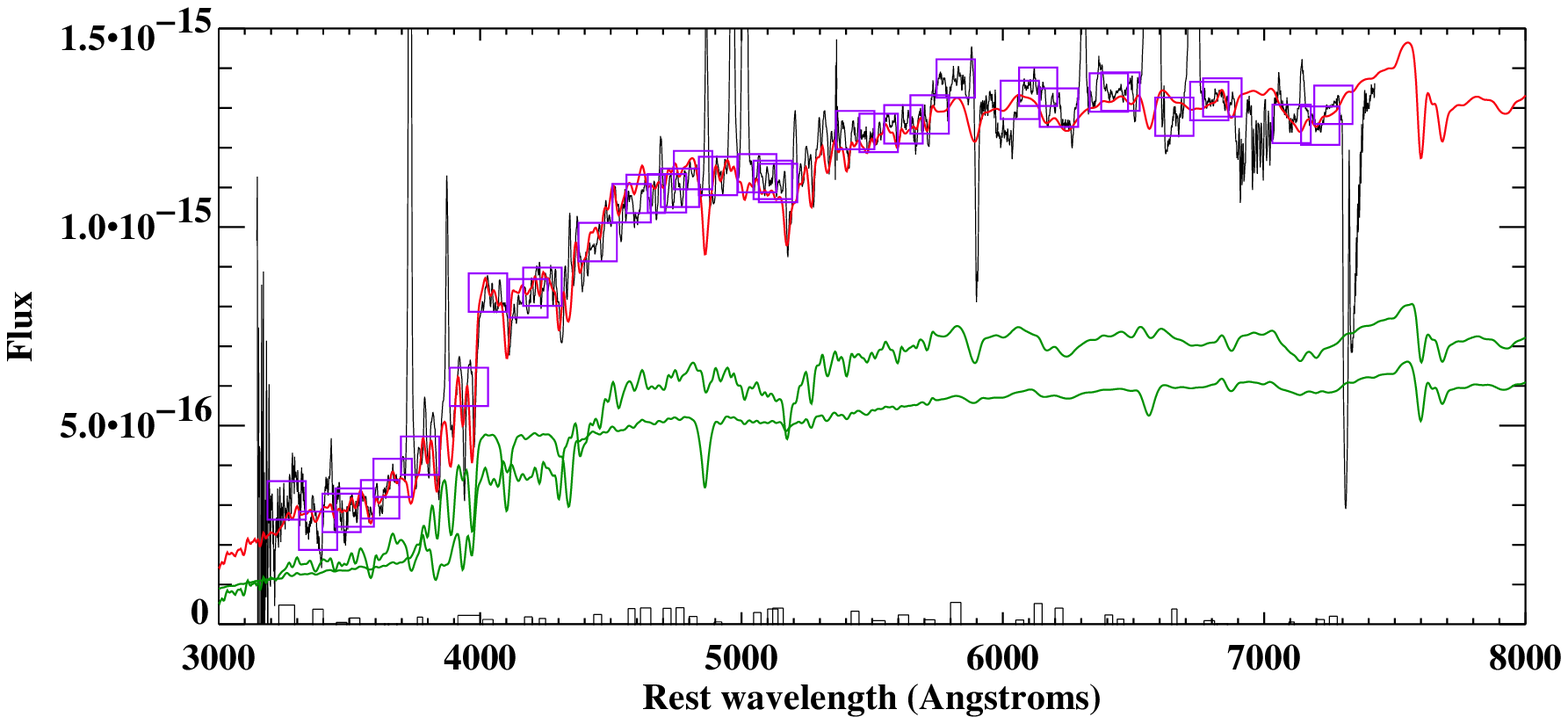,width=13.0cm} \\
\psfig{figure=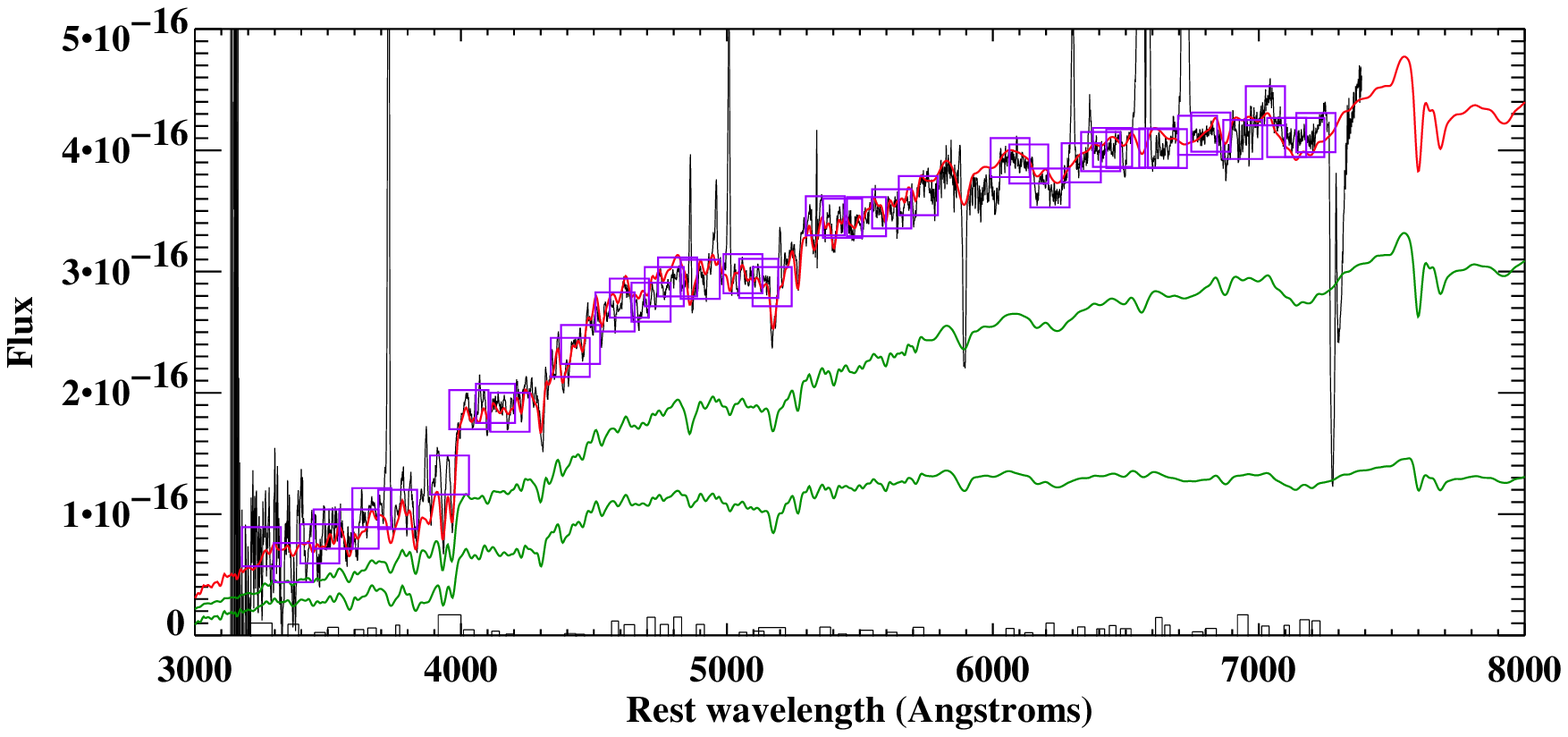,width=13.0cm} \\
\vspace{0.2cm}\\
\end{tabular}
\caption{\label{fig: examples}. Examples of model fits to the nuclear spectra of 3C305(top) and 3C293 
(bottom),
showing the best fitting models surperimposed on the data, the data bins, the components to the fits, and the 
absolute residuals of the data from the models (the latter as rectangular columns at the bottom of the plot). In 
the case of 3C305 the best fitting model comprises a 12.5Gyr elliptical (54\%) 
plus a 0.4Gyr YSP reddened by
E(B-V)$=$0.6 (46\%), while in the
case of 3C293 the best fitting model comprises a 12.5Gyr elliptical (38\%) plus a 2.0Gyr 
YSP reddened by E(B-V)$=$0.4 (62\%), where the percentages refer to the contributions to
the total model flux in the normalising bin. Further details of the fitting procedure are given in section 4.2.}
\end{figure*}

All three objects in the sample show a significant excess of UV flux,
when compared with a passively evolving old elliptical; values of
\( D(4000) \) suggest that only \( 25 \)-- \( 50 \) \% of the flux
in the UV is due to the \( 12.5 \) Gyr stellar population. In all
cases, the excess is spatially extended, and only the nucleus of 3C~305 shows 
a value of \( D(4000) \) consistent with an old stellar
population. Tadhunter et al. (2002) found \( D(4000) \)
in the range \( \sim 1.2 \) -- \( 2.6 \) for a large sample of 2Jy radio galaxies  in
the redshift range ($0.02<z<0.7$). Our measured values of \( D(4000) \)
are somewhat to the lower end of this range, suggesting a large contribution from YSP or activity-related
continuum on radial scales of 1 --- 20~kpc.

Various spatial regions of interest were identified in each object
using the continuum spatial profiles. These are marked in Figure
1. The spectra of these spatial regions were
then extracted from the 2-d frames and used in the continuum modelling
process detailed below. The extracted spectra for the blue/green spectral
range are shown in Figures 2 \& 3.

\subsection{Continuum modelling}

Following the clear detection of UV excesses in all the sample objects, 
the nature of the optical/UV continua were
investigated by modelling the detailed shapes of the
continuum spectral energy distributions 
(SEDs: see Tadhunter et al. 1996, Robinson et al. 2000). 
Note that we use SED modelling rather than absorption line indices
for this work, because most of the age-sensitive diagnostic absorption 
lines are strongly affected by emission line infilling 
(CaII~K is an exception --- see section 4.3 below).   
Initially, a nebular continuum was generated for each of the spatial
regions, comprising the blended higher ($>$H8) Balmer series, together
with a theoretical nebular continuum (a combination of free-free
emission, free-bound recombination and two-photon continua) generated
using the Starlink software DIPSO. This was then subtracted from the
data prior to the modelling (see Dickson et al 1995). For some of
the data (notably the nucleus of PKS 1345+12), H\( \beta  \) has
a multiple-Gaussian line profile; in this case separate nebular continua
were generated with the same redshift and line width as each emission
component. For all but the nuclear region in PKS1345+12 the contribution of 
nebular continuum to the UV continuum was found to be relatively small ($<$10\%).

After subtraction of the nebular continuum, four sets of models of increasing 
complexity were
fitted to the data. As before, all the stellar
population models used in the fitting assume a Salpeter IMF, solar-metallicity instantaneous
starburst. First, a model comprising a 12.5~Gyr elliptical
template taken from BC96 (see section 4.1) was fitted to the data. Second, models were constructed from two
components: a power-law of the form
$F_{\lambda }\propto \lambda ^{\alpha }$, taken to represent either direct or scattered AGN light, and
a 12.5~Gyr-old elliptical galaxy template. Third, a series of models
was constructed that comprise a 12.5~Gyr elliptical and a contribution from a young stellar
population (YSP), with the age of the YSP chosen to be in
the range 0.01 --- 2.5~Gyr. Finally, a set of models that combine an old
elliptical, a YSP and a power-law was used. 

For the nuclei of 3C 293 \& 3C 305, the models with unreddened
YSP did
not give satisfactory fits to the data (see below). In these cases,
a further series of models was generated, comprising an old elliptical
(12.5 Gyr) combined with a reddened YSP. In the latter case, YSPs
with $0 < E(B-V) < 1.6$ were considered. The Seaton (1979) reddening law
was used to redden the YSP spectra.

Continuum bins were selected from a list of possible rest frame bins,
chosen to avoid emission lines and atmospheric absorption bands. Typically 40 -- 45 bins were
selected for each object, with the bins chosen to be as evenly distributed in wavelength as
possible, in order to avoid biasing the fits towards a particular wavelength region. A
normalising bin, common to all three objects was chosen with the wavelength
range 4700 --- 4730\AA.  The models were then generated by scaling the
different components so that the total model flux in the normalising
bin was $<$125\% of the observed flux\footnote{The maximum allowed model flux is greater 
than 100\% of the measured flux to allow for
uncertainties in the measured flux and models in the bin.}. The best-fitting model with
each set of components, and for each age of YSP, was determined using
a chi-squared minimisation technique over the model and data continuum
bins. To ensure reasonable calculation times, the power-law spectral index was
limited to the range ($-15 < \alpha <15 $). 

For the chi-squared 
fitting we assumed an error of $\pm$5\% in each wavelength
bin --- consistent with the relative flux calibration error. 
We find that the flux calibration error dominates over photon counting and sky subtraction
errors. Note that, since the flux calibration errors are unlikely to be normally distributed, and are not
independent between the data bins, we have not assigned formal confidence intervals on the
basis of the reduced chi-squared values for the fits; the reduced chi-squared values provide
an indication of the region of parameter space for the models that provides a good fit to the data, 
rather than accurate statistical uncertainties. However, it is clear from a visual
inspection of the fits that models with $\chi^2_{red} < 1$ give acceptable fits, whereas
those with  $\chi^2_{red} > 1$ do not provide an adequate despription of the spectral shape
our high quality data, in the sense that the measured fluxes in several adjacent wavelength bins
differ systematically from the model by more than 5\% of the total flux. This suggests that the 
assumed $\pm$5\% errors are realistic. 
On this basis we regard any model with $\chi^2_{red} < 1$ as acceptable.

As examples of our fitting procedure, we show fits to the nuclear spectra of 3C305 and 3C293 in Figure
4, with the continuum bins and residuals from the fits clearly indicated. 

We consider the results of the model fits for each object in the following
sections.

\subsubsection{3C293}

Example SED modelling results for 3C293 are presented in Figures 4, 5 \& 6  and Table 2.

\begin{table*}
{\centering \begin{tabular}{ccccccc}
{\par}&
OE \%&
YSP \%&
YSP age (Gyr)&
PL \%&
PL \( \alpha  \)&
\( \chi ^{2} \)\\
\hline
 3C293 E1&
\( 54\pm 4 \)&
\( 54\pm 5 \)&
\( 1 \)&
---&
---&
\( 0.65 \)\\
&
\( 38_{-13}^{+14} \)&
\( 59_{-15}^{+13} \)&
\( 1 \)&
\( 8\pm 7 \)&
\( 2.3_{-1.4}^{+6.1} \)&
\( 0.33 \)\\
\hline
3C293 E2&
\( 41\pm 4 \)&
\( 64\pm 5 \)&
\( 1 \)&
---&
---&
\( 0.50 \)\\
&
\( 33_{-14}^{+16} \)&
\( 64_{-15}^{+14} \)&
\( 1 \)&
\( 7\pm 7 \)&
\( 1.5_{-7.8}^{+13.4} \)&
\( 0.37 \)\\
\hline 
3C 293 N&
\( 82\pm 10 \)&
\( 6_{-6}^{+7} \)&
\( 0.5 \)&
\( 17_{-7}^{+6} \)&
\( 3.0_{-0.6}^{+1.1} \)&
\( 0.47 \)\\
\hline
3C305 E1&
\( 28\pm 4 \)&
\( 77\pm 5 \)&
\( 1 \)&
---&
---&
\( 0.70 \)\\
&
\( 16_{-14}^{+17} \)&
\( 79_{-18}^{+15} \)&
\( 1 \)&
\( 8\pm 8 \)&
\( 1.9_{-2.2}^{+13.0} \)&
\( 0.43 \)\\
\hline
3C305 E2&
\( 91\pm 3 \)&
\( 17\pm 3 \)&
\( 0.5 \)&
---&
---&
\( 1.31 \)\\
&
\( 73\pm 10 \)&
\( 25_{-7}^{+6} \)&
\( 0.5 \)&
\( 5_{-5}^{+6} \)&
\( 3.8_{-1.3}^{+3.7} \)&
\( 0.66 \)\\
\hline
PKS1345+12 E1&
\( 21\pm 4 \)&
\( 86\pm 5 \)&
\( 1 \)&
---&
---&
\( 0.54 \)\\
&
\( 21_{-14}^{+5} \)&
\( 74_{-12}^{+13} \)&
\( 1 \)&
\( 4_{-4}^{+6} \)&
$\pm$15&
\( 0.52 \)\\
\hline
\end{tabular}\par}

\caption{Best-fitting models to the continuum in selected regions
of 3C 293, 3C305 and PKS1345+12 (unreddened YSP only). Percentages are quoted relative 
to the observed flux in the range 4700 -- 4730 \AA.}
\end{table*} 

For the two extended apertures in 3C293 (E1 and E2) we find that the optical SEDs can only
be fitted adequately if we include a YSP component that  contributes between  20 and 100\%
of the light in the normalising aperture (depending on the age of the YSP). Inclusion of a
power-law component as well as a YSP improves the fits marginally. However, the
contribution of the power-law component in the normalising bin is relatively minor, and the
power-law slope relatively red, suggesting that the power-law may be compensating for a
small amount of reddening and/or inadequacies in the flux calibration.

\begin{figure}
\begin{tabular}{c}
\psfig{figure=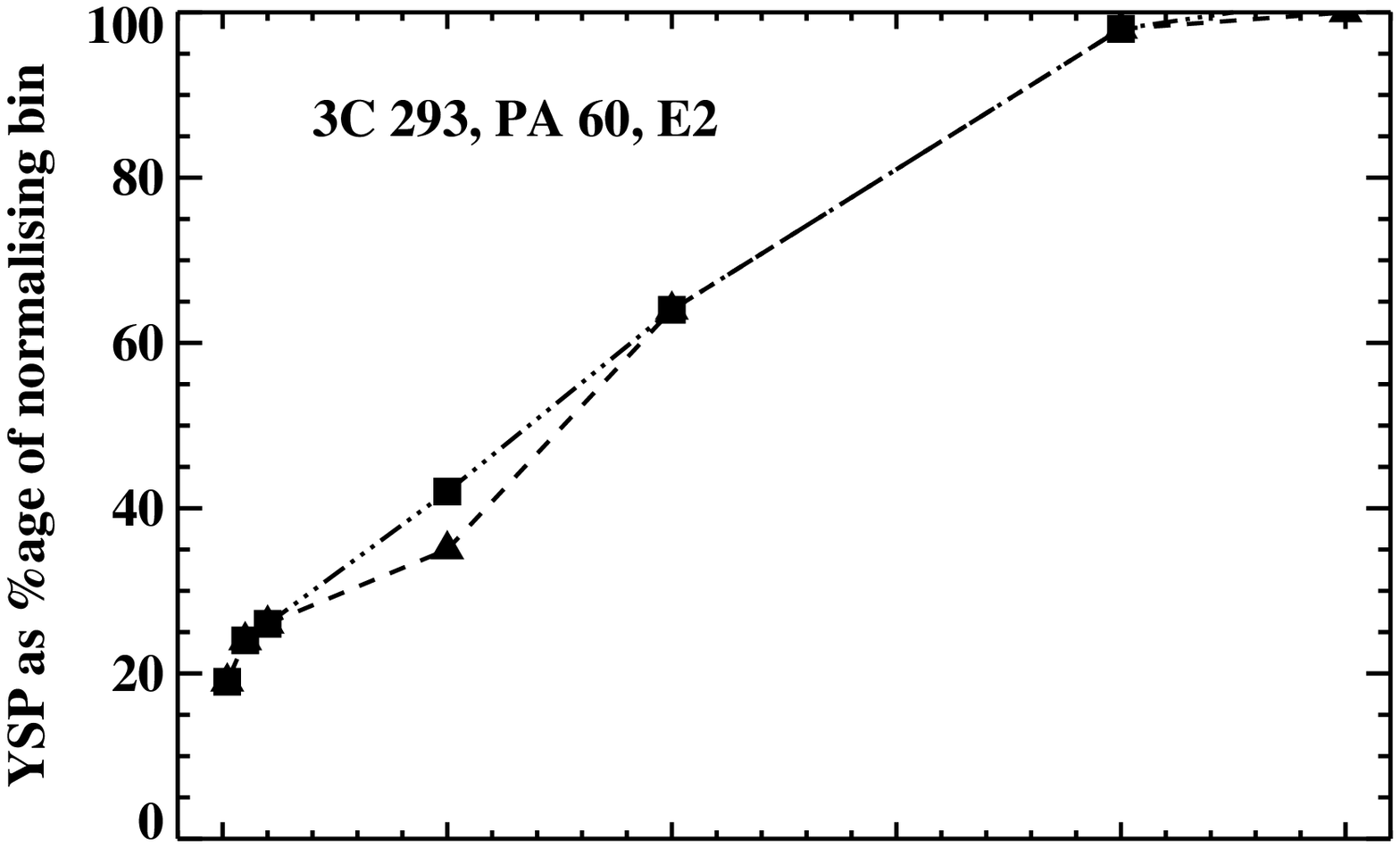,width=8.0cm} \\
\psfig{figure=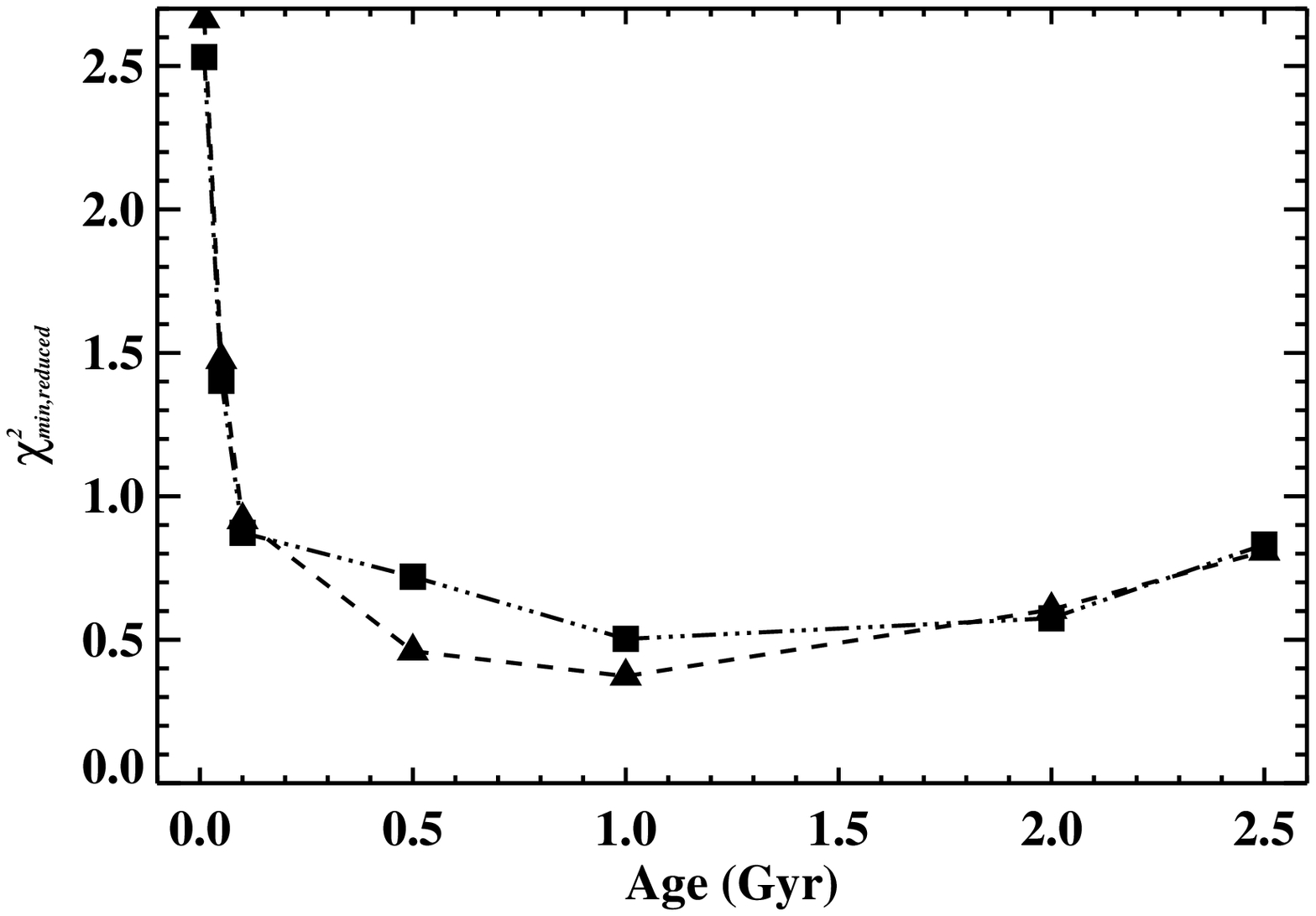,width=8.0cm} \\
\vspace{0.2cm}\\
\end{tabular}
\caption{\label{fig: 293 E2 results}Results of continuum modelling for 3C
293, E2. (Top) Percentage of YSP in best-fitting model at that age
vs age of YSP. (Bottom) Value of \protect\( \chi _{min,reduced}^{2}\protect \)
vs. age. If \protect\( \chi ^{2}>1\protect \), fits are not significant.
Dot-dash lines denote models without power-law components, dashed
lines are models that include a power-law. The points on each line
reflect those ages at which models were generated.}
\end{figure}

The ages of the YSP in the extended regions of this object are not well-constrained by the 
SED modelling alone. We find that, although the SEDs are best fit with YSP ages of
$\sim$1~Gyr, all YSP ages greater than 0.075~Gyr produce an acceptable fits to the SEDs
(Figure 5). We
face the problem that the results are degenerate. For example, the SED modelling does not
distinguish between, on the one hand  a continuum dominated by an old stellar population
plus a moderate contribution from a relatively young YSP ($<0.3$~Gyr), 
and on the other a
single intermediate age ($\sim$2.0 ­-- 2.5~Gyr) stellar population.

\begin{figure}
\begin{tabular}{l}
\psfig{figure=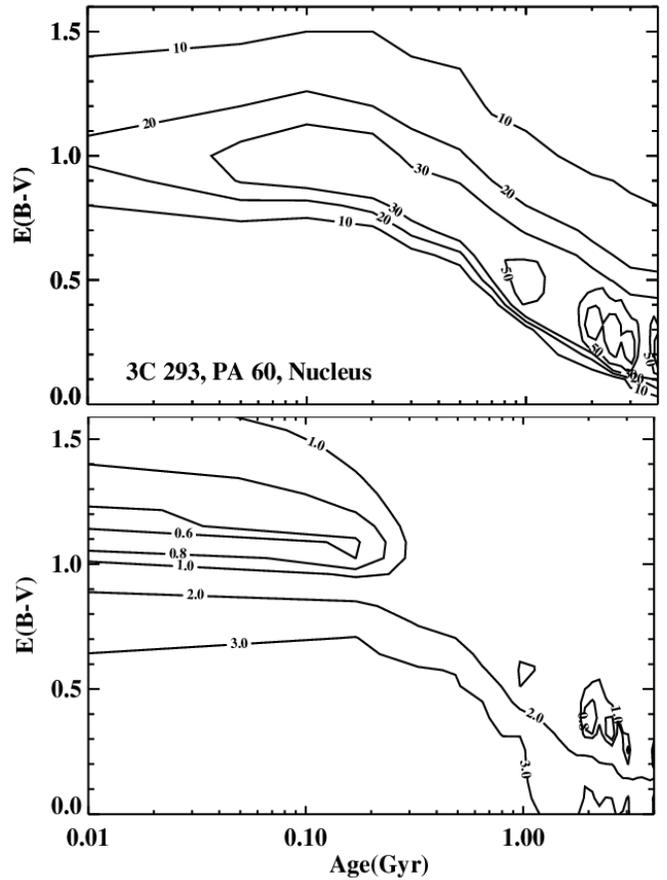,width=9.0cm} \\
\vspace{0.2cm}\\
\end{tabular}
\caption{\label{fig: 293 nucleus - reddening contour plots}Results of fitting
reddened YSPs to 3C 293, nucleus. (Top) Contour plot representing
the percentage contribution of the YSP for the best-fitting model
at a given age and reddening. (Bottom) Contour plot representing the
value of \protect\( \chi _{min,reduced}^{2}\protect \) for the best-fitting
model. In both cases, no power-law component is fitted.}
\end{figure}

In the nuclear regions of 3C293, despite the tentative detection of absorption underlying the
Balmer emission lines, a UV excess, and star clusters detected in UV HST images (Allen et al. 2002), the
overall continuum SED is relatively red (see Figure 4). Consequently, it is not possible to obtain an
acceptable fit to the nuclear SED with models comprising solely an old elliptical plus
unreddened YSP; acceptable fits can only be obtained with the inclusion of a significant red
($\alpha \sim 3$) power-law component. However, HST imaging results presented by Chiaberge et al. (1999)
show no evidence for
a power-law point source component in 3C293, despite the detection of point sources in many of the
objects in their sample with fluxes of 
$\sim1\times10^{-17}$ erg cm$^{-2}$ s$^{-1}$ \AA$^{-1}$ at $\sim$7000 \AA\, --- much smaller than the
putative power-law component required by our SED modelling.
The redness of the required power-law and 
the presence of dust lanes in HST images of the nuclear regions suggests an alternative
model that comprises a combination of a reddened YSP and an old elliptical galaxy component. The 
contours of reduced $\chi^2$ shown in Figure 6 allow us to determine which
combinations of reddening (parameterised as E(B-V)) and YSP age provide an adequate fit 
to the nuclear continuum. We find that acceptable fits are obtained with an elliptical
galaxy component plus either a relatively young YSP ($<0.3$~Gyr) with a large reddening 
($E(B-V) > 1.0$), or a relatively old YSP ($1.0$ ­-- $2.5$~Gyr) with moderate reddening ($0.3 <
E(B-V) < 0.5$).

Therefore, despite the clear presence of YSP in all three apertures modelled for this source, 
we find that we cannot determine the ages, relative contributions and reddening of the YSP
based on the SED modelling alone. In section 4.3 below we discuss a method for breaking 
this degeneracy, based on the strength of the CaII~K absorption line.

\subsubsection{3C305}

Example SED modelling results for 3C305 are presented in Figures 4, 7 \& 8 and Table 2.

Some aspects of the modelling results for 3C305 are similar to those for 3C293. For example, in the
extended regions of 3C305 we find that we can obtain a reasonable fit with a model comprising an 
old elliptical plus a YSP, and can improve the fits with the addition of a red-power law that
contributes a relatively small proportion of the flux in the normalising bin. In the extended 
regions of this  object, however, the ages are better constrained than in 3C293, with YSP
ages in the range 0.4 ­-- 1.5~Gyr favoured by the models, and YSP ages younger than 
0.1~Gyr and older than 2.0~Gyr ruled out (Figure 7).

\begin{figure}
\begin{tabular}{c}
\psfig{figure=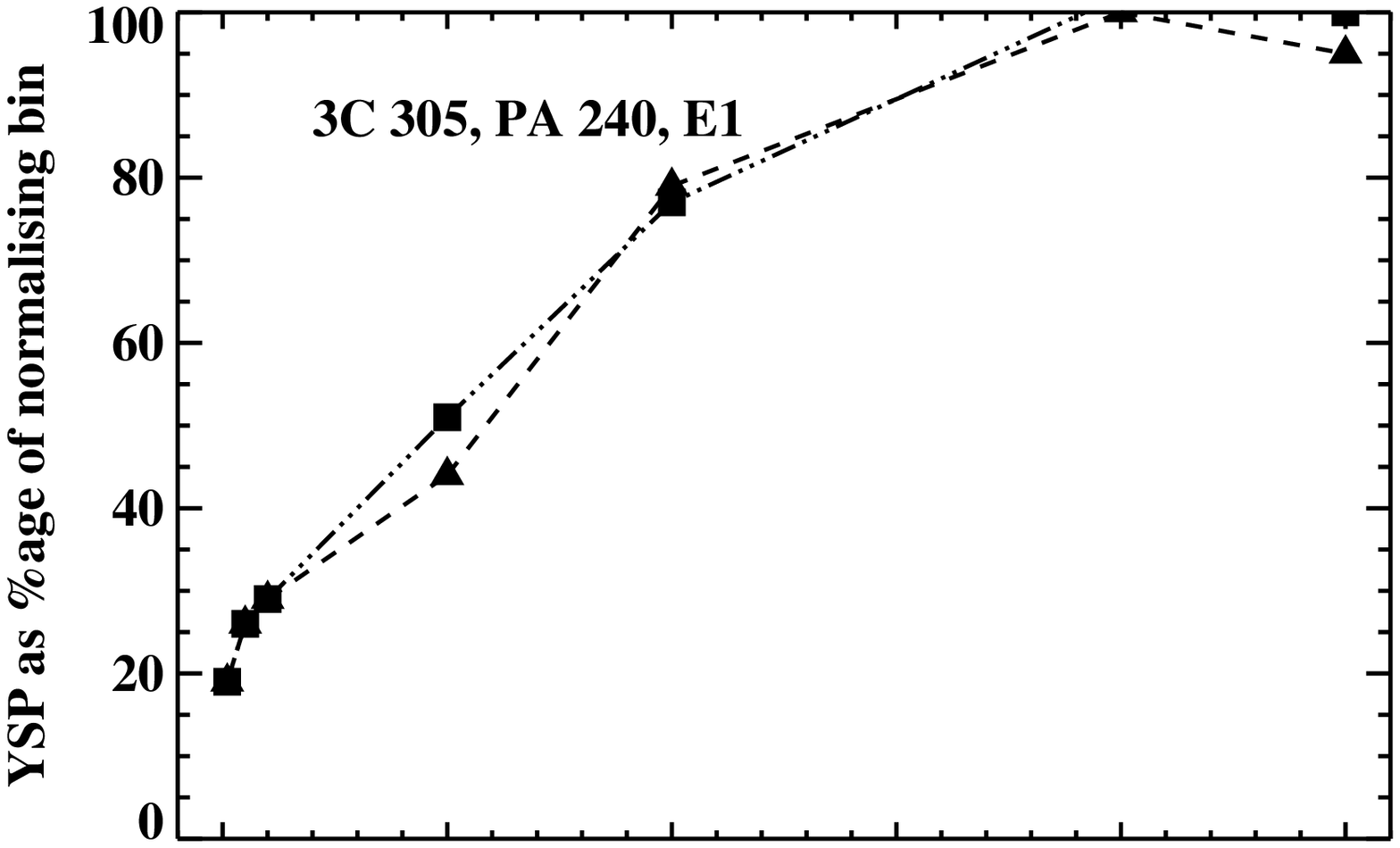,width=8.0cm} \\
\psfig{figure=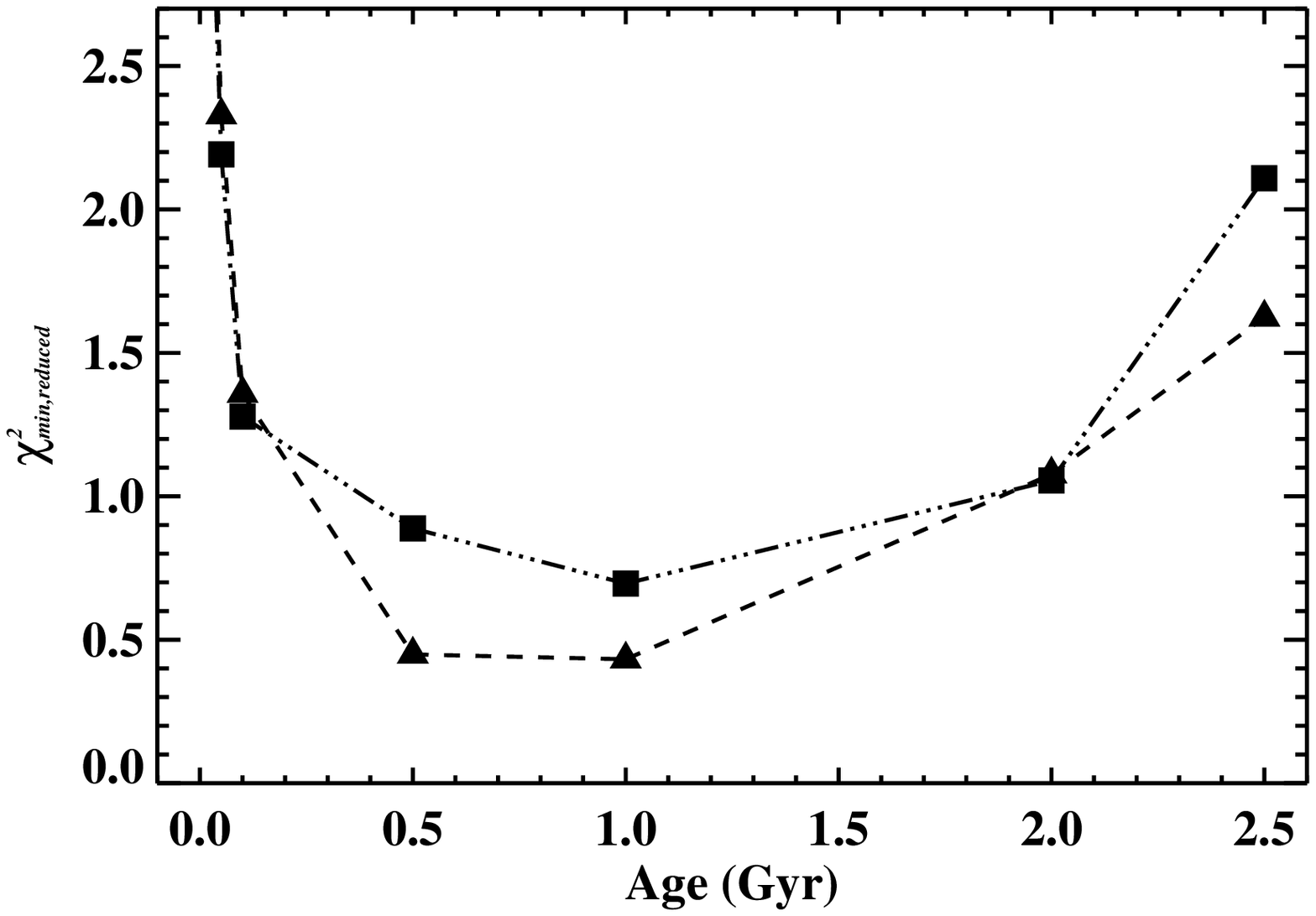,width=8.0cm} \\
\vspace{0.2cm}\\
\end{tabular}
\caption{\label{fig: 305 e1 results} Results of continuum modelling for 3C
305, E1. (Top) Percentage of YSP in best-fitting model at that age
vs age of YSP. (Bottom) Value of \protect\( \chi _{min,reduced}^{2}\protect \)
vs. age. If \protect\( \chi ^{2}>1\protect \), fits are not significant.
Dot-dash lines denote models without power-law components, dashed
lines are models that include a power-law. The points on each line
reflect those ages at which models were generated.}
\end{figure}

In the case of the nuclear aperture, we find that no combination of old stellar 
population, unreddened YSP and power-law, or any of these two components alone can
provide an adequate fit to the nuclear spectrum in this source. However, a combination of a 
reddened YSP plus an old stellar population provides an excellent fit to the nuclear SED for
YSP ages in the range  $0.2 < t_{YSP} < 1.0$~Gyr and reddening in the range $0.4 < E(B-V) < 
0.8$, with a best fit obtained for  $t_{YSP} \sim 0.4$~Gyr and $E(B-V) \sim 0.6$ (Figure 8). Note that the
age deduced for the YSP in the nuclear regions is consistent with those deduced for the 
extended regions in this source, and that this object does not suffer from the degeneracy
problem noted for 3C293. Nonetheless, in section 4.3 below we will check the results of 
the SED modelling using CaII~K line strength measurements.

\begin{figure}
\begin{tabular}{c}
\psfig{figure=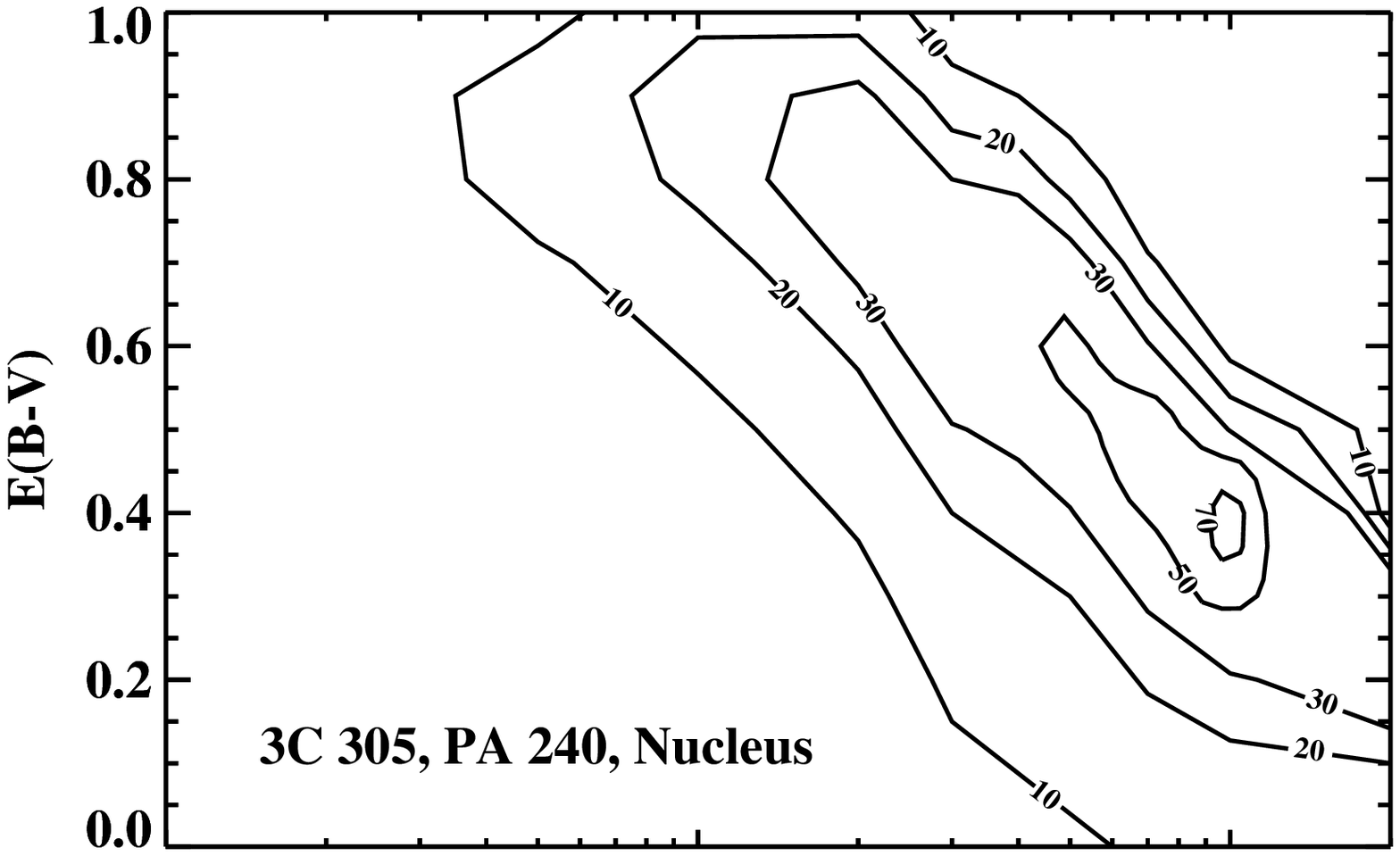,width=8.0cm} \\
\psfig{figure=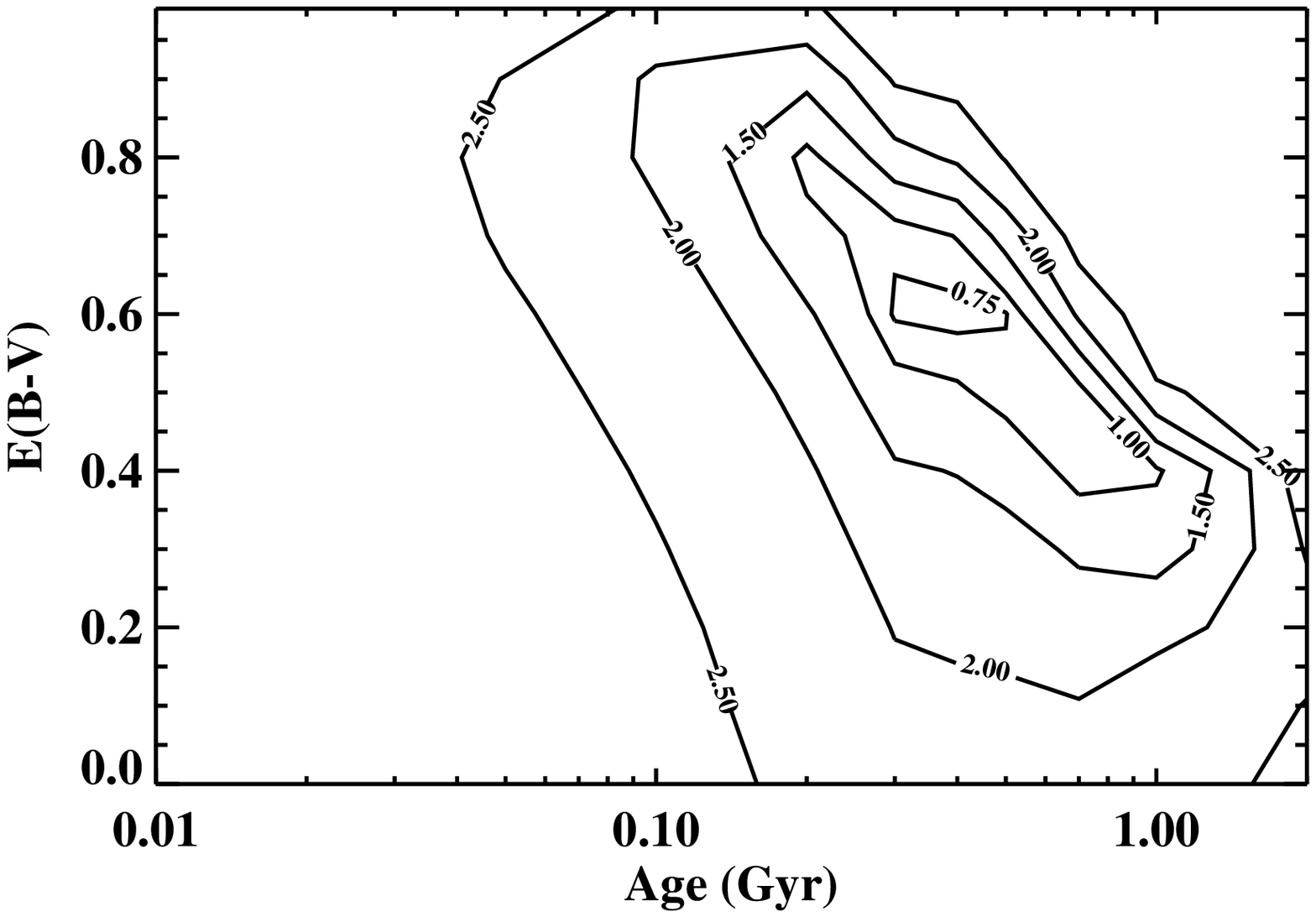,width=8.0cm} \\
\vspace{0.2cm}\\
\end{tabular}
\caption{\label{fig: 305 nucleus - reddening contour plots}Results of fitting
reddened YSPs to 3C 305, nucleus. (Top) Contour plot representing
the percentage contribution of the YSP for the best-fitting model
at a given age and reddening. (Bottom) Contour plot representing the
value of \protect\( \chi _{min,reduced}^{2}\protect \) for the best-fitting
model. In both cases, no power-law component is fitted.}
\end{figure}

\subsubsection{PKS1345+12}

Example SED modelling results for PKS1345+12 are presented in Figure 9 and Table 2.

The nuclear regions of PKS1345+12 are extremely active, with luminous, highly
reddened emission line regions that are undergoing outflow from the nucleus (Holt et al. 
2003). Because of the reddening of the emission line regions, it is difficult to subtract the 
nebular
continuum adequately in the nuclear aperture. Moreover, the strength and breadth of the 
emission lines from the 
outflowing gas make it difficult to identify sufficient continuum bins for the SED
modelling. Overall, the extreme activity precludes accurate stellar population analysis close 
to the nucleus of this source. Therefore we have concentrated on the extended region
centred 5 arcseconds to the north west of the nucleus (E1). In this region we find that we can 
obtain an accurate fit to the SED with a combination of an old elliptical galaxy plus an
unreddened YSP with age in the range $0.5 <  t_{YSP} < 1.5$~Gyr (Figure 9). The addition of a power-
law component does not significantly improve the fits. Note that this is the first direct
spectroscopic detection of a YSP in this source. We check the age of the YSP using the 
CaII K line strength in the next section.

\begin{figure}
\begin{tabular}{c}
\psfig{figure=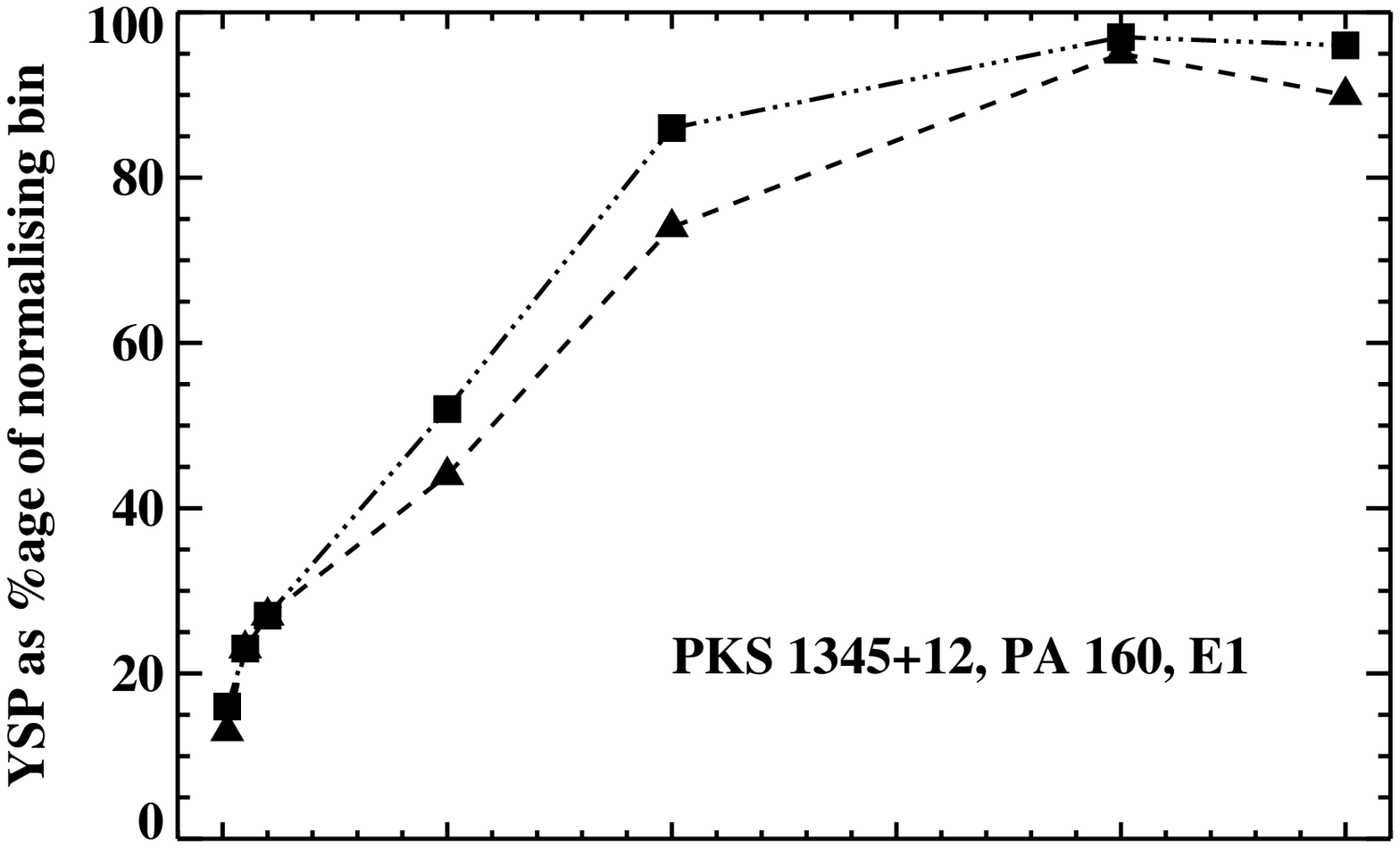,width=8.0cm} \\
\psfig{figure=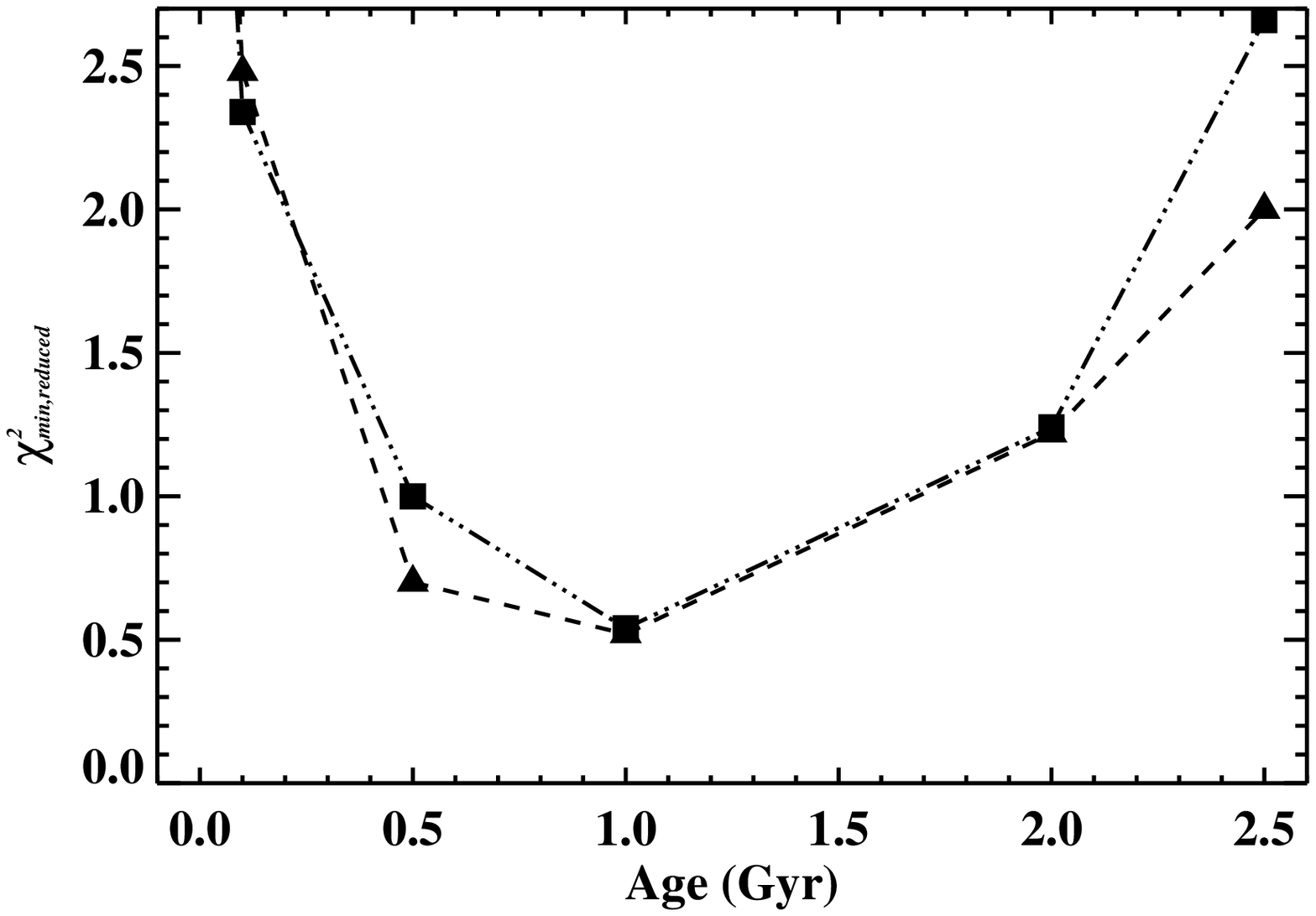,width=8.0cm} \\
\vspace{0.2cm}\\
\end{tabular}
\caption{\label{fig: 1345 e1}As Figure 5, but for
PKS 1345+12, E1.}
\end{figure}

\subsection{Line strengths}

The SED modelling described above is our primary analysis tool. However, the 
stellar absorption line features provide useful
complementary information which allows us to check the SED modelling results and resolve 
ambiguities.

First we note that 
H\( \delta  \) and higher order Balmer lines are detected in absorption 
in several spatial regions of our objects (see Figure 2), despite the potential for emission line
contamination. Since these features are too weak
to detect in stellar populations with ages above 5~Gyr, 
their detection in the radio galaxies provides unambiguous evidence for
a significant contribution from young or intermediate age stellar populations
(Gonz\'alez Delgado, Leitherer \& Heckman 1999). This supports one of the major 
conclusions of our SED modelling.

Useful information on the young stellar populations can also be obtained using
the age-sensitive CaII~K line, which is not significantly affected by emission line contamination.
This line has the potential of allowing us to break the degeneracy between the age and relative
contribution/reddening of the YSP in the spectral synthesis results that was noted above.

Since the resolution of the BC96 models is too low for detailed modelling of 
absorption features, we have compared our spectra instead with the higher resolution spectral synthesis
results of Gonz\'alez Delgado  et al. (1999, 2001) over a limited spectral range surrounding the Ca II~K line.
In doing so, we have used the results of the SED modelling from section 4.2 as a guide to the 
relative age
and contribution of the YSP. Because we only consider a narrow spectral range we have not taken
into account reddening of the YSP component. The results are shown in Figures 10, 11 \& 12.

First we consider 3C293, for which either an elliptical plus a moderate contribution from a relatively
young YSP, or an elliptical plus a large contribution from an intermediate age YSP, provide good 
fits to the SED. In the case of the nuclear aperture of 3C293 we find that the
strength of the Ca II~K line is underpredicted by the E+young YSP model, however, the E+intermediate age
YSP model provides an excellent fit to this line (see Figure 10). Therefore, based on both the SED fitting and the
CaII~K line strength, we conclude that the age of the YSP required to fit the nuclear spectrum
in 3C293 is 1.0 --- 2.5~Gyr; in this case we have broken the degeneracy between the age and the
reddening/proportional contribution of the YSP. Similar results are obtained for aperture E2, however, 
in region E1 we find
evidence for significantly younger ages, as signalled by the relatively weak CaII~K line
(see Figure 10, top two plots). In this
latter region, the combined SED and CaII~K results suggest YSP ages in the range 0.1 --- 0.5~Gyr.
Note that this model also provides an excellent fit to the H$\delta$ line in this region, which is
relatively free of emission line contamination.

In the case of 3C305 (see Figure 11), the CaII K results are broadly consistent with the SED
modelling, but favour the upper end of the YSP age range derived from the SEDs 
(0.6 --- 1.0~Gyr in the case of the nuclear aperture); models with YSP ages $t_{YSP} < 0.4$~Gyr 
underpredict the CaII K line strength.

Finally, in the case of PKS1345+12 E1, we find that the CaII K results are consistent with
the YSP ages derived from the SED modelling ($0.5 < t_{YSP} < 1.5$~Gyr), and the model shown in
Figure 12, which comprises a 60\% contribution from a 0.7Gyr YSP and a 40\% contribution from
an elliptical, provides an
excellent fit to the CaII~K line.

\begin{figure*}
\begin{tabular}{c}
\psfig{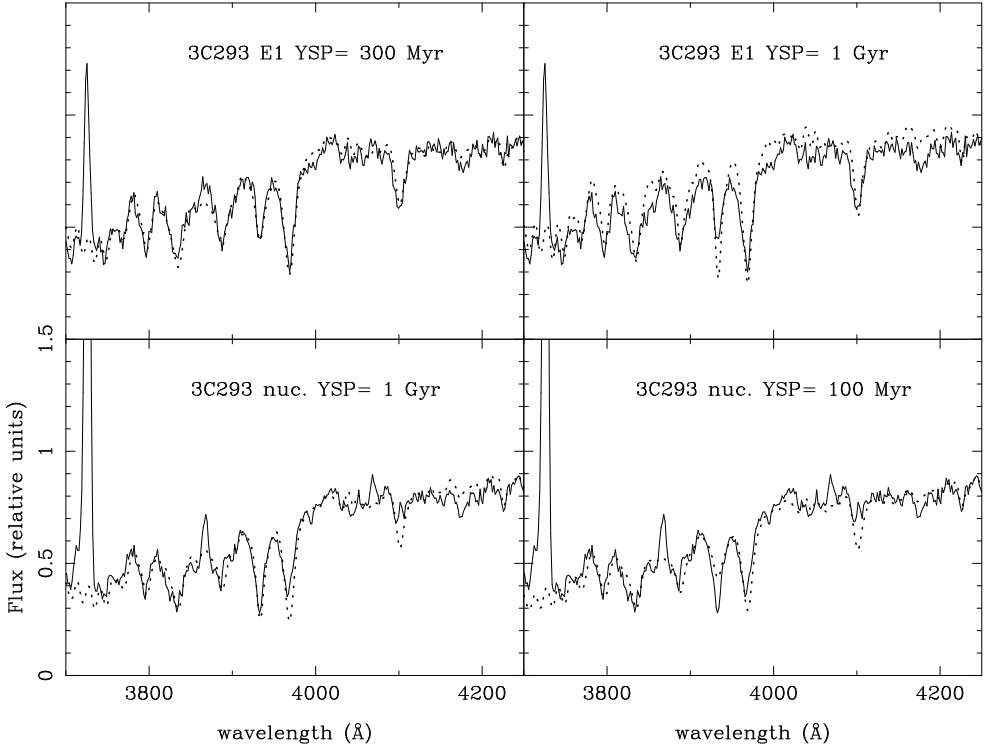}
\end{tabular}
\caption{Detailed fits to the nuclear (bottom) and E1 (top) spectra of 3C293 for different YSP contributions and ages.
In each case the solid line shows the spectrum of the source, and the dotted line the model fit.
The different models are as follows. Lower left: nuclear spectrum with 50\% 1~Gyr YSP plus 
50\% elliptical model. Lower right: nuclear spectrum with 35\% 0.1~Gyr YSP plus 65\% elliptical model.
Upper left: E1 spectrum with 30\% 0.3~Gyr YSP plus 70\% elliptical model. Upper right: E1 spectrum with
70\% 1~Gyr YSP plus 30\% elliptical model. The relative proportions of the different components are
determined in a normalising bin at 4700\AA. The key CaII~K feature is at a rest wavelength of 3934\AA.}
\end{figure*}

\begin{figure*}
\begin{tabular}{c}
\psfig{figure= 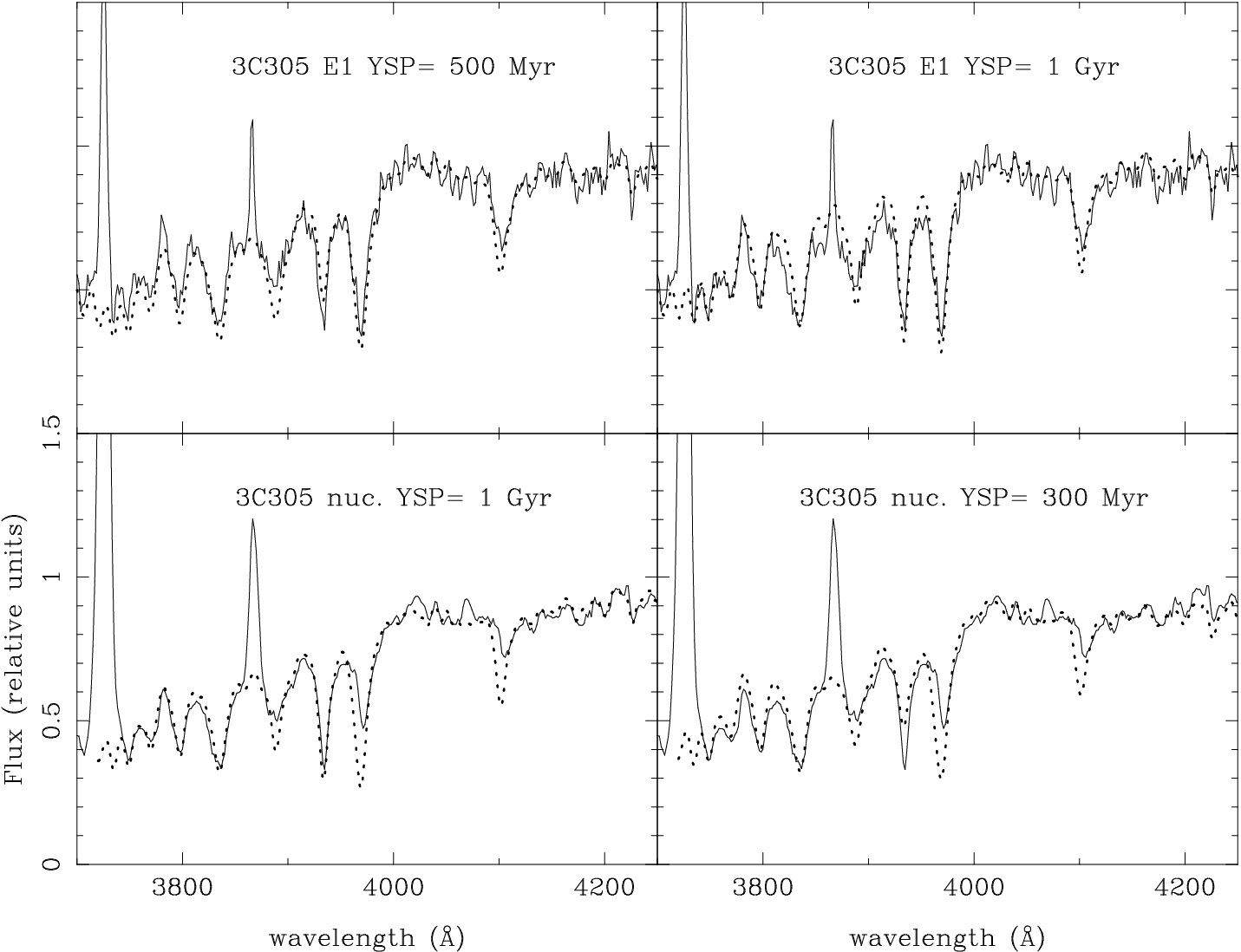,width=13.0cm,angle=-90}
\end{tabular}
\caption{Detailed fits to the nuclear (bottom) and E1 (top) spectra of 3C305 for different YSP contributions and ages.
In each case the solid line shows the spectrum of the source, and the dotted line the model fit.
The different models are as follows. Lower left: nuclear spectrum with 70\% 1~Gyr YSP plus 
30\% elliptical model. Lower right: nuclear spectrum with 35\% 0.3~Gyr YSP plus 65\% elliptical model.
Upper left: E1 spectrum with 50\% 0.5~Gyr YSP plus 50\% elliptical model. Upper right: E1 spectrum with
20\% 1~Gyr YSP plus 80\% elliptical model. The relative proprtions of the different components are
determined in a normalising bin at 4700\AA.}
\end{figure*}

\begin{table*}
{\centering \begin{tabular}{ccclll}
&
&
Age of &
\( 12.5 \) Gyr &
YSP&
YSP \% \\
&
&
YSP &
mass &
mass&
of total\\
&
&
/ Gyr&
/ \( M_{\odot } \)&
/ \( M_{\odot } \)&
mass\\
\hline
3C 293
&E1
&0.1 -- 0.5 
&(1.0 -- 2.6)$\times10^{10}$
&(0.50 -- 2.2)$\times10^8$
&0.4 -- 4.6 \\
&E2
&1.0 -- 2.5
&(0.0 -- 4.9)$\times10^9$
&(0.67 -- 3.4)$\times10^9$
&12 -- 100
\\
&N
&1.0 -- 2.5
&1.0$\times10^{10}$
&(0.80 -- 1.1)$\times10^{10}$
&44 -- 52
\\
\hline
3C 305
&E1
&0.5 -- 1.5
&(0.85 -- 4.1)$\times10^{9}$
&(0.88 -- 1.9)$\times10^{9}$
&31 -- 51
\\
&E2
&0.5 -- 1.0
&(1.3 -- 1.5)$\times10^{10}$
&(1.3 -- 4.1)$\times10^{8}$
&0.9 -- 3.0\\
&N
&0.4 -- 1.0
&(2.6 -- 5.6)$\times10^{10}$
&(1.1 -- 1.9)$\times10^{10}$
&16 -- 42\\
\hline
PKS 1345 +12
&E1
&0.5 -- 1.5
&(0.28 -- 1.7)$\times10^{10}$
&(0.77 -- 3.7)$\times10^{9}$
&4 -- 56
\\
\hline
\end{tabular}\par}

\caption{Ages and masses for the young and old stellar populations for the various spatial regions of 
the objects. In each case a range of values is given  to cover the range of models that give an acceptable
fit to the SEDs and CaII~K absorption lines. The final column gives the percentage of the total stellar mass
contributed by the YSP in the regions covered by the slit.}
\end{table*}

\subsection{Summary of key results}

\begin{table*}
{\centering \begin{tabular}{clll}
&
Bolometric YSP &
Absorbed &
Measured far-IR \\
Object&
luminosity / \( L_{\odot } \)&
luminosity / \( L_{\odot } \)&
luminosity / \( L_{\odot } \)
\\
\hline
3C293
&(0.67 -- 1.0)$\times 10^{10}$
&(4.3 -- 8.7)$\times 10^{9}$
&1.5$\times10^{10}$
\\
3C305
&(2.4 -- 2.6)$\times 10^{10}$
&(1.8 -- 2.3)$\times 10^{10}$
&1.3$\times10^{10}$
\end{tabular}\par}

\caption{YSP bolometric luminosities calculated over
the wavelength range \protect\( 0\protect \) -- \protect\( 30,000\protect \)
\AA. The range corresponds to the range of models that provide an acceptable fit to the spectra. 
Also shown (third column) are the luminosities absorbed or scattered by dust
in the nuclear regions, calculated from the amount of
reddening required in the best-fitting models. The final column shows
the (\protect\( 60\protect \) -- \protect\( 100\protect \)
\protect\( \mu \protect \)m) far-IR luminosities, calculated from the
IRAS results of Golombek et al. (1988).}
\end{table*}

A summary of the properties of the YSP derived from this study is presented in Table 3. 
As well as our best estimates of the ages, we provide estimates of the masses of YSP in the 
regions covered by the slits, and the proportion of the total in-slit stellar mass contributed 
by the YSP. For the nuclear regions the YSP masses were estimated following correction 
for reddening. All the estimates of ages and masses are based on the spectral synthesis 
results of BC96 for an instantaneous starburst,  Salpeter IMF with mass range 
0.1 ­-- 125~M$_{\odot}$ and solar metallicity. 

Since we are also interested in the relationship between the YSP and far-IR luminosities 
in the sample objects, we have estimated the bolometric luminosities for the nuclear YSP, and the amount of far-IR
radiation absorbed by the extinguishing dust, for comparison with the measured far-IR luminosities
of the targets. The results are shown in Table 4. Using the spectral synthesis models we can also ``turn the
clock back'' and estimate the bolometric luminosities that the nuclear YSP would have had in the past. 
The results -- which are based on the instantaneous burst models of
BC96 -- are shown in graphical form in Figure 13 for 3C305 and 3C293, where they are compared
with the luminosities of luminous- and ultra-luminous infrared galaxies (LIGs and ULIGs: see 
Sanders \& Mirabel 1996 for definition).  

The key results from our study are as follows.
\begin{itemize}
\item{\bf Ages of the YSP.} The oldest YSP components detected in all three galaxies have 
post-starburst ages in the range $0.5 < t_{YSP} < 2.5$~Gyr, although there is evidence for 
a younger YSP in region E1 of 3C293.
\item{\bf Reddening.} Unsurprisingly, given the imaging evidence for dust lanes, the YSP 
detected in the near-nuclear regions of 3C293 and 3C305 are significantly reddened.
\item{\bf Masses of the YSP.} The masses of the YSP are substantial, particularly in the 
nuclear regions of 3C293 and 3C305 where they are comparable to masses of molecular gas estimated on the 
basis of measurements of CO in the core regions of 3C293 and PKS1345+12 (Evans et al. 1999a,b). 
\item{\bf YSP contribution to the total stellar mass.} The YSP contribute a substantial proportion 
of the total in-slit stellar mass in all the regions they are detected, even in the near-nuclear regions 
where the contribution from the old stellar populations in the galaxy bulges might be 
expected to be substantial.
\item{\bf Current far-IR luminosties.} For 3C305 and 3C293 the YSP luminosities absorbed/scattered 
by dust are similar to the current far-IR luminosities of the galaxies. This supports the idea that 
the dust radiating the far-IR light is heated by the post-starburst components detected in our
spectra. 
\item{\bf Past far-IR luminosities} On the basis of the  
results presented in Figure 13, and assuming
complete absorption and reprocessing of the radiant luminosity of the YSP by dust, 
it is clear that
3C293 and 3C305 would have appeared as ULIGs if observed within 10~Myr of the start of the starbursts. This
assumes that the starbursts were instantaneous, however, even for non-instantaneous starbursts, it is clear that
these objects would have appeared as LIGs up to 0.1~Gyr after the start of the starbursts, provided that
the star formation episode did not last longer than 0.1~Gyr. 
\end{itemize}

\subsection{Uncertainties in the derived properties}

A major advantage of our SED modelling technique compared with the alternative method 
of using the absorption line indices is that it allows us to account directly for all the 
activity-related and stellar components that contribute to the observed spectra. Moreover, because 
of their wide spectral coverage, our WHT/ISIS data are sensitive to a range of stellar 
population ages, reddened or unreddened. The complementary information provided by the 
CaII~K line allows us to refine age estimates based on the initial SED modelling. Therefore 
we are confident about our modelling technique and the suitability of our dataset for studies 
of this type.

Most probably, the largest sources of uncertainty in the results described above relate to the assumed 
parameters of the spectral synthesis results used to model the data, most notably, the 
duration of the starburst, the abundances, and the IMF shape and mass range.  

In the above analysis we assumed that the starburst was instantaneous in each region of 
each galaxy. In reality, the star formation history might have been more complex, with 
multiple bursts, or continuous star formation. The effect that more complex star formation 
histories would have on our results can be gauged by considering the extreme case in which, stars have 
actually formed at a continuous rate, but that we have attempted to model the SED with an 
instantaneous burst YSP model. A continuous, or quasi continuous, star 
formation will always produce an SED that is bluer, has a larger UV excess and a smaller 
Balmer break than an instantaneous burst model of similar age. Therefore, attempting to 
model such an SED with an instantaneous burst model will always underestimate the time 
since the start of the star formation episode, and the ages of YSP estimated above will be  
lower limits on the time since the start of the starbursts. This strengthens the conclusion 
that the post-starburst YSP in radio galaxies are relatively old.

Considering the issue of the abundances,  on the basis of our analysis and the 
morphological study of Heckman et al. (1986) it is likely that the starbursts have been 
triggered by the interaction between two or more galaxies, at least one of which is a gas-rich 
disk galaxy; we can rule out the idea that the activity has been triggered by a minor 
accretion event such as, for example, the accretion of a dwarf irregular galaxy. Thus we expect that 
the abundances of the gas out of which the stars form in the starburst are likely to be 
moderately enriched i.e. close to solar; it is unlikely that our results are affected by 
abundances that are significantly sub- or super-solar.

Perhaps the most serious uncertainty, however, is the exact shape and mass range of the 
initial mass function (IMF).  An IMF that in 
reality is steeper than Salpeter, or has a relatively large low mass cut-off, would 
lead us to overestimate the total mass of the starburst component.  For example, if the 
lower mass cut-off of the IMF was at 1.0M$_{\odot}$ rather than 0.1M$_{\odot}$, or the 
IMF had a steeper, Scalo power-law index (-3.3), then the assumptions we have used above 
will lead us to overestimate the total mass in the YSP by a factor $\sim$2 ­-- 3. The study of 
the IMF shape for clusters in nearby merger systems is in its infancy. However, despite the 
controversial result of Smith \& Gallagher (2001) for the cluster M82-F, recent results for the
Antennae merger system are  
consistent with either a Salpeter or Scalo IMF with lower mass cut-off at 0.1M$_{\odot}$, 
or a Salpeter IMF with lower mass cut-off at $\sim$1.0 (Mengel et al. 2002). 
Therefore, the results in Table 3 are 
unlikely to be uncertain by more than a factor of $\sim$2 ­-- 3.

Overall, despite uncertainties related to the assumptions of spectral synthesis 
modelling procedure, we believe that the general results listed in section 4.4 are likely 
to hold.

\section{Discussion}

\subsection{The nature of the young stellar populations}

In the following we will assume that the YSP represent post-starburst stellar populations, 
and that the starbursts have been triggered by galaxy mergers or interactions. However, alternative 
possibilities  include the capturing of the YSP associated with the disks 
of one or more of the merging galaxies, and/or the jet-induced star formation associated with the activity. 

In the case of captured stellar population, the ages of the YSP reflect the 
star formation histories of the disks prior to the mergers, and cannot be used to estimate the 
time scales and order-of-events in the mergers that trigger the activity. 
Although we cannot entirely discount this possibility, it seems 
unlikely given that: (a) the masses of the YSP are large; (b) much of the star formation is 
concentrated within a few kpc of the nuclei of the host galaxies; and (c) the ages of the YSP 
in the extended regions are similar to those in the nuclear regions. If all the YSP were 
captured in these objects, then it is unlikely that they would be so strongly concentrated in 
the nuclear regions. On the other hand, circum-nuclear starbursts are predicted by models 
for gas-rich mergers, since tidal torques associated with the mergers concentrate gas in the 
near-nuclear regions. Indeed, CO observations provide direct evidence for large gas 
concentrations in the nuclear regions of 3C293 and PKS1345+12 (Evans et al. 1999a,b).

In the case of jet-induced star formation, there are three aspects that make this scenario unlikely. First, the ages
of the YSP are significantly larger than the expected ages of the high 
surface brightness radio sources (see section 5.4 below). Second, the published optical, UV and radio images show no 
clear morphological  association or alignment between radio and optical/UV continuum features. Finally, the YSP
extend well beyond the maximum extent of the high surface brightness radio structures for the slit PAs used for
the observations of all three objects.
The only possibility that we cannot entirely exclude is that the YSP are associated with past cycles of radio
activity. However, give the general lack of evidence for jet-induced star formation in the
population of radio galaxies, we regard this possibility as unlikely.   

We conclude that the YSP in the nuclear regions of 3C293 and 3C305 represent post-starburst 
stellar populations, and that the YSP in the extended regions of these two galaxies 
are also likely to represent post-starburst populations since their ages are similar to those in 
the nuclear regions. PKS1345+12 is a less certain case because we have not yet 
been able to compare the age of its circum-nuclear starburst with that of the extended YSP detected
in our observations. However, even if the YSP in the extended region of PKS1345+12 are captured, this is still
consistent with the idea that we are witnessing a major merger in which one of the galaxies has a substantial
disk component.    

\begin{figure}
\begin{tabular}{c}
\psfig{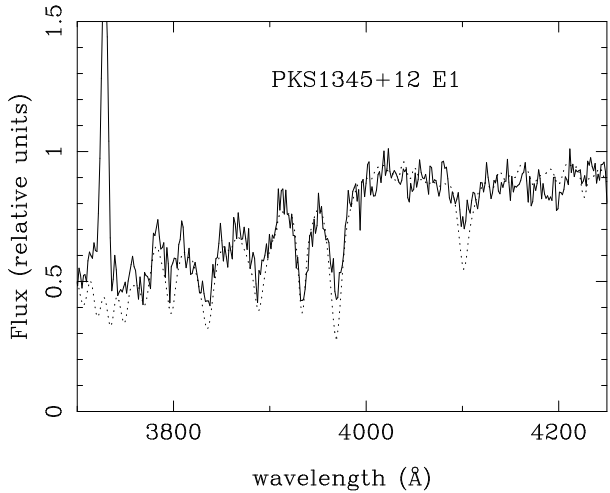}
\end{tabular}
\caption{Detailed fit to the e1 spectrum of PKS1345+12 for a 60\% 0.7~Gyr YSP plus elliptical model.
The relative proprtions of the different components are
determined in a normalising bin at 4700\AA.}
\end{figure}

\subsection{Evolutionary status}

The large proportional contribution of the YSP to the total stellar masses of the host 
galaxies implies that we are witnessing the aftermath of major mergers between massive
galaxies, at least one of which must have been gas-rich. It is likely that these major mergers 
have triggered both the starbursts and 
the AGN activity, although not necessarily 
at the same time  (see section 5.4 
below).

It is interesting to consider these results in the context of the evolution of the population of 
early-type galaxies. Hierarchical galaxy evolution models suggest that a subset of the 
elliptical galaxy population was formed and/or has undergone substantial evolution via 
major mergers at relatively recent epochs (e.g. Kauffmann, Haehnelt \& White 1996). Observationally, 
the recent evolution of the 
elliptical galaxy population is supported by measurement of line-strength indices in some 
early-type galaxies that imply the presence of significant YSP (see Kuntschner 2000
and references therein). In addition, infrared studies 
of ULIGs demonstrate that at least some remnants of 
major mergers have kinematic properties similar to those of early-type galaxies (Genzel et al. 2001). 
On the 
other hand, the interpretation of the redshift evolution of the number density of early-type 
galaxies is more controversial (e.g. Zepf 1997, Cimatti et al. 2002, Rodighiero et al. 2002), 
and it is unlikely that {\it all} early-type galaxies have undergone 
major evolution in recent past. Indeed, the evidence suggests that both the formation epoch 
and speed of the evolution of early-type galaxies depend on environment (e.g. Menanteau,
Abraham \& Ellis 2002). Whereas much 
of the observational evidence for the recent evolution of the early-type galaxy population is 
confined to the field, or relatively low luminosity E/SO galaxies on the periphery of 
clusters (Kuntschner 2000), giant elliptical galaxies 
in rich clusters show little evidence for recent evolution.

Therefore it is plausible that the three radio galaxies discussed in this paper are associated 
with the subset of the early-type galaxy population in relatively low density environments 
that is undergoing the most rapid evolution via mergers in the local universe. The fact that 
none of the three galaxies is in a rich cluster environment is consistent with this hypothesis.

\subsection{A link between ULIGs and radio galaxies?}

The status of the radio galaxies in our sample as merging systems raises the issue of 
whether all merging systems that end up as early-type galaxies go through a radio galaxy 
phase as part of their evolution. Therefore it is interesting
to investigate the links between the radio galaxies, luminous infrared galaxies
(LIGs: $L_{ir} > 10^{11}$ L$_{\odot}$) and 
ultraluminous infrared galaxies (ULIGs: $L_{ir} > 10^{12}$ L$_{\odot}$) ­-- some 
of the most extreme merging systems in the local universe (see Sanders \& Mirabel 1996 for
a review). The evidence for such links
includes the 
following.
\begin{itemize}
\item[-]{\bf Optical morphologies.} In terms of their optical morphologies there are clear similarities 
between radio galaxies and ULIGs. The three galaxies in our sample form part of the  
the subset of $\sim$50\% of all nearby powerful radio galaxies that show evidence for morphological 
features such as tidal tails, arcs, bridges, and double nuclei that are characteristic of mergers
(Heckman et al. 1986). 
These features have a higher surface brightness than the faint shell structures commonly 
detected in nearby ``normal'' elliptical galaxies. Therefore, in terms of the degree of 
morphological disturbance, radio galaxies fall between ULIGs, which show such high surface brightness features 
in $>$95\% of cases (Sanders \& Mirabel 1996), and normal elliptical galaxies in which such features are much rarer.   

\item[-]{\bf Far-infrared luminosities}. Only $\sim$30\% of powerful radio galaxies at low redshifts $z < 0.3$ were 
detected at far-IR wavelengths by the IRAS satellite (Golombek et al. 1988, Impey \& Gregorini 1993, 
Heckman et al. 1994); and, apart from a few rare exceptions, the far-IR 
luminosities of the detected sources are well below those of ULIGs. This implies that, if 
a substantial proportion of ULIGs go through a radio galaxy phase, the radio galaxy 
phase cannot be coeval with the ULIG phase in all objects. It is notable
that, in line with previous results 
which suggest a strong link between optical/UV starburst activity and far-IR luminosity in 
radio galaxies (Tadhunter et al. 2002, Wills et al. 2002), all three of the objects in our 
sample were detected by IRAS, albeit covering 
a wide range in far-IR luminosity. Whereas PKS1345+12 is one of the few radio galaxies 
in the local universe classified as a ULIG based on its far-IR luminosity, the other two 
sources ­-- 3C293 and 3C305 ­-- have far-IR luminosities that are orders of magnitude lower. 
However, we showed in section 4.4 that, assuming instantaneous starbursts and complete repreocessing 
of the optical/UV light by dust, both of these objects would have appeared as LIGs or ULIGs in the past.
Even 
allowing for the fact that the starbursts may not be instantaneous, these results provide clear 
evidence for a link between radio galaxies and luminous- or ultraluminous infrared galaxies, 
in the sense that some LIGs or ULIGs may evolve into radio galaxies. This direction of evolution 
is supported by the fact that the ages of the post-starburst populations in the three objects in 
our sample are older than those of all the ULIG quasar host galaxies investigated using 
similar techniques by Canalizo \& Stockton (2001).  

\item[-]{\bf Space densities.} By integrating the $z=0$ radio luminosity
function of Dunlop \& Peacock (1990) we find that the space density of extragalactic radio sources in the
local universe with radio powers greater than or equal to that of 3C305 ($P_{2.7GHz} > 5\times10^{25}$W Hz$^{-1}$)
--- the least luminous radio source
in our sample --  is $\rho(PRG) = 8.5\times10^{-7}$ Mpc$^{-3}$ for our assumed cosmology . For comparison  
the space densities of LIGs and ULIGs,
found by integrating the far-infrared luminosity function of Sanders \& Mirabel (1996), are
$\rho(LIG) = 2.7\times10^{-5}$ Mpc$^{-3}$ and $\rho(ULIG) = 1.2\times10^{-7}$ Mpc$^{-3}$ respectively. 
It is notable that
the space density of moderately powerful radio galaxies is $\sim$7$\times$ greater than that of the ULIGs,
but $\sim$30$\times$ less than that of LIGs.We also find that the space density of the far-IR sources becomes
equivalent to that of the moderately luminous radio sources for $L_{ir} > 4\times10^{11}$ L$_{\odot}$.
Even allowing for the fact that not all radio sources may be triggered by major
galaxy mergers (see Tadhunter et al. 1989 and  Baum et al. 1990 for a discussion), these results are 
consistent with the idea that all ULIGs and a significant fraction of the LIGs
evolve into moderately powerful radio sources. On the other hand, assuming that
the timescale of the ULIG/LIG phase is not substantially less than the timescale of the radio source phase,
the fact that the space density of radio
sources is much larger than that of ULIGs suggests that not all moderately luminous radio sources were ULIGs
in the past, although many of them may have been LIGs. 

\item[-]{\bf Stellar kinematics.} Based on optical spectroscopic data, Heckman et al. (1986) and Smith,
Heckman \& Illingworth (1990) showed that nearby powerful radio galaxies with strong 
emission lines have stellar kinematic properties similar to those of ``disky'' E-galaxies, 
with relatively large $V_c/\sigma$ providing evidence for rotational support. These kinematic 
properties are identical to those deduced from recent near-IR spectroscopic observations that 
penetrate the dusty environments of the bulges of ULIG galaxies (Genzel et al. 2002); the mean velocity dispersions
measured for the ULIGs studied by Genzel et al. (2001: ($<\sigma> = 185\pm 14$) are similar
to those measured for the radio galaxies with strong emission lines by Smith et al.(1990: $<\sigma> = 209\pm 13$).
\end{itemize}

%
%

Overall, several lines of evidence are consistent with the idea that ULIGs and LIGs can evolve into radio
galaxies. Further studies of the stellar populations in ULIGs, LIGs and radio galaxies would help to put 
this putative evolutionary sequence onto a firmer footing.    

\subsection{Triggering and order-of-events}

An intriguing feature of the results obtained above is that the post-starburst stellar 
populations in all of our sample objects are relatively old ($\sim$0.5 ­-- 2.5~Gyr) compared 
with the lifetimes of the radio sources estimated from standard spectral ageing arguments 
($<$0.1~Gyr). Taken at face value this result implies that the radio sources are 
triggered relatively late in the merger sequence -- much later than the starbursts. This ties
in with results based on HI~21cm and CO imaging observations, which also suggest a substantial
delay between the start of the merger and the triggering of the jets 
(Evans et al. 1999a,b, Morganti et al. 2003). 

What causes the apparent time lag between the starburst and jet triggering events? One possibility is 
that the lag represents a settling timescale: the time taken for the 
gas to establish a steady accretion flow into the nuclear regions after it has been stirred up 
in a major circum-nuclear starburst. Alternatively, the lag may represent the time required to 
form the type of AGN that is capable of powering powerful relativistic jets. For example, it 
has been suggested that powerful radio jets are associated with rapidly spinning black holes 
formed by the coalescence of the black holes in the nuclei of two galaxies undergoing a 
major merger (Wilson \& Colbert 1995). In this case, the starburst/radio source time lag may be
related to the 
coalescence timescale for the binary black hole. 

Before leaving this topic it is important to add a caveat about the radio source ages. 
Recently, the standard spectral ageing estimates of radio source lifetimes have been 
challenged by Blundell \& Rawlings (2001) who argue that radio lobes may be continually 
replenished by energetic electrons, so that the high frequency breaks/curvature in the radio 
spectrum may not reflect the ages of the radio sources. In this case, the most extended 
double radio sources may be much older (0.1 ­-- 1~Gyr) than predicted by traditional spectral 
ageing arguments. However, two of the sources in our sample are compact radio sources 
and, for such sources, spectral ageing arguments are thought to provide accurate radio 
source lifetimes because they agree with dynamical age estimates. Therefore, at least for the 
two compact objects in our sample, it is difficult to escape the conclusion that there is a 
substantial time lag between the starburst that produced the YSP we now observe at optical 
wavelengths, and the triggering of the currently observed radio jets. 

Another possibility is that the radio sources 
have undergone multiple epochs of activity, with one epoch coincident with the original 
starburst. Indeed, the mutiple epochs idea is supported by the detection of low surface 
brightness extended radio emission surrounding the much higher surface brightness compact 
radio sources in some objects, including PKS1345+12 (Stanghellini et al. 2001). Moreover,
the ``double-double'' radio structure of 3C293,
may also suggest multiple epochs of activity. However, if radio jets are always 
triggered at the same time as the original merger-induced starbursts, observations
of larger samples of radio galaxies should reveal the presence of YSP with
relatively young ages ($<$0.1~Gyr) in at least some sources.

\begin{figure}
\psfig{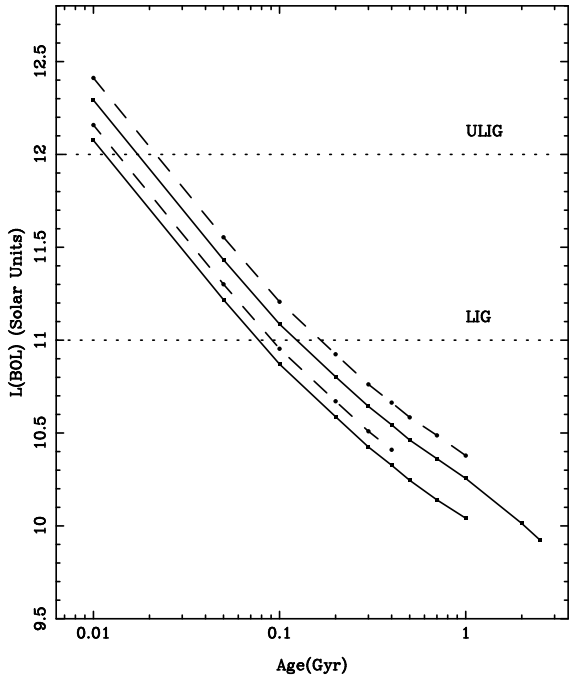}
\caption{The evolution of the bolometric luminosities of the YSP components with age for the nuclear region
of 3C305 (dashed lines), and the nuclear region plus region E2 of 3C293 (solid lines), as predicted by the BC96 instantaneous burst models. For each object two lines are shown,
corresponding to the maximum and minimum age YSP that give acceptable fits to both the SEDs and
the CaII~K 
absorption line. For reference, the limiting luminosities for LIGs and ULIGs are also shown
(see Sanders \& Mirabel 1996). }
\end{figure} 

\section{Conclusions and further work}

The results presented in this paper demonstrate the potential of studies of the young stellar 
populations  for investigating the nature of the events that trigger the activity in radio 
galaxies, the order of events, the timescales, and the relationship between radio galaxies and
other types of active objects such as ULIGs. The main results can be summarised as 
follows.
\begin{itemize}
\item{\bf Nature of the YSP and the mergers.} The YSP are identified with post-starburst populations 
that are relatively  old and massive, and represent a large fraction of the total stellar mass in 
the regions sampled by our spectra. These results imply that we are observing the aftermath 
of {\it major mergers} between massive galaxies, at least one of which must have been a gas-rich
disk galaxy.

\item{\bf Spatial distribution.} Although the YSP are concentrated close to the nuclei of the host 
galaxies in at least two of our sample objects, in all cases the YSP are spatially extended on 
a scale of 1 ­-- 20~kpc.

\item{\bf Triggering and timescales.} The fact that the post-starburst populations detected in our 
sources are relatively old compared with the typical ages of extragalactic radio sources, suggests that
the activity associated with the observed high surface brightness radio sources was triggered late in the merger 
sequence --- up to 0.5 -- 2~Gyr after the starburst responsible for the YSP.

\item{\bf Luminous and ultraluminous infrared galaxies.} Our results point to a strong link between the radio galaxies and at least a subset of the
LIG and ULIG populations; they provide evidence that some LIGs and ULIGs may {\it evolve into} radio galaxies.
\end{itemize}

It is now important to extend such studies to larger samples of radio galaxies, in order to put 
these results on a firmer footing, and investigate whether they
hold for the general population of radio galaxies with detected YSP.

\subsection*{Acknowledgments} TR and KW acknowledge support from PPARC. RGD and KW acknowledges support from the
Royal Society and by the Spanish Ministry of Science and Technology (MCyT) through grant 
AYA-2001-3939-C03-01. We acknowledge useful comments from the anonymous referee.

{}

\end{document}